\documentclass[a4paper]{cas-dc}

\usepackage[numbers]{natbib}
\usepackage{amsmath,amsthm,amssymb,amsfonts}
\usepackage{multirow}
\usepackage{makecell}
\usepackage{caption}
\usepackage{color}
\usepackage{soul}
\usepackage{hyperref}
\newcommand{\onlinecite}[1]{\hspace{-1 ex} \nocite{#1}\citenum{#1}}
\newcommand{\bt}{\textbf}

\begin{document}
\let\WriteBookmarks\relax
\def\floatpagepagefraction{1}
\def\textpagefraction{.001}
\let\printorcid\relax
% Short title
\shorttitle{} 
\shortauthors{Lingxiao Lei et al.} 

% Main title of the paper
\title [mode = title]{Cryogenic Control and Readout Integrated Circuits for Solid-State Quantum Computing} 

%%%%%%% first author %%%%%%%
\author[2]{Lingxiao Lei}
\credit{Conceptualization of this study, methodology, software}

\author[1]{Heng Huang}
\credit{simulation, supervision, writing}

\author[2]{Pingxing Chen}
\credit{supervision, writing}

%%%%%%% second author %%%%%%%
\author[1]{Mingtang Deng}
\ead{mtdeng@nudt.edu.cn}
\credit{supervision, writing}
\cormark[1]
\cortext[1]{Corresponding author.}
%\orcidlink{0000-0000-0000-0000}

%%%%%%% affiliations %%%%%%%
\affiliation[1]{organization={Institute for Quantum Information \& State Key Laboratory of High Performance Computing, College of Computer Science and Technology, National University of Defense Technology},
city={Changsha},
postcode={410073}, 
country={China}}
\affiliation[2]{organization={Institute for Quantum Science and Technology, College of Sciences, National University of Defense Technology},
city={Changsha},
postcode={410073}, 
country={China}}

\begin{abstract}
In the pursuit of quantum computing, solid-state quantum systems, particularly superconducting ones, have made remarkable advancements over the past two decades. However, achieving fault-tolerant quantum computing for next-generation applications necessitates the integration of several million qubits, which presents significant challenges in terms of interconnection complexity and latency that are currently unsolvable with state-of-the-art room-temperature control and readout electronics. Recently, cryogenic integrated circuits (ICs), including CMOS radio-frequency ICs and rapid-single-flux-quantum-logic ICs, have emerged as potential alternatives to room-temperature electronics. Unlike their room-temperature counterparts, these ICs are deployed within cryostats to enhance scalability by reducing the number and length of transmission lines. Additionally, operating at cryogenic temperatures can suppress electronic noise and improve qubit control fidelity. However, for CMOS ICs specifically, circuit design uncertainties arise due to a lack of reliable models for cryogenic field effect transistors as well as issues related to severe fickle noises and power dissipation at cryogenic temperatures. This paper provides a comprehensive review of recent research on both types of cryogenic control and readout ICs but primarily focuses on the more mature CMOS technology. The discussion encompasses principles underlying control and readout techniques employed in cryogenic CMOS ICs along with their architectural designs; characterization and modeling approaches for field effect transistors under cryogenic conditions; as well as fundamental concepts pertaining to rapid single flux quantum circuits. In conclusion, while prototypes of cryogenic ICs have been successfully fabricated recently, further studies are still necessary; we believe that rapid single-flux-quantum-based or hybrid ICs featuring enhanced digital components may represent the future direction in this emerging area.

\end{abstract}


\maketitle

\section{Introduction}\label{Intro}

%\bt{\blue{Introduction to QC}}. 
Quantum computing, as an emerging computing paradigm in the post-Moore era, could leverage distinctive properties of quantum mechanics such as superposition and entanglement. Consequently, it offers exponential reductions in computational complexity for specific problems that pose challenges to classical electronic computers, including factoring large integers~\cite{ShorSIAM1999}, database searching~\cite{GroverPRL1997}, and simulating quantum dynamics~\cite{Georgescu_Rev_Mod_Phys_2014}.

%\bt{\blue{Introduction to Solid-State QC, restrict discussion scope}}. 
In recent decades, a wide range of quantum systems have become controllable in laboratory settings. These systems encompass ions~\cite{BruzewiczAPR2019}, photons~\cite{Kok_Rev_Mod_Phys_2007, BrienScience2007}, neutral atoms\cite{HenrietQuantum2020, wurtz2023aquila}, and others. Among them, solid-state platforms such as the Josephson-junction-based superconducting platform and semiconductor quantum dot platforms stand out due to their scalability and well-established control technology. The scalability and control capabilities of these platforms are facilitated by micro/nano processing techniques and microwave engineering~\cite{Bardin_Journal_of_Microwaves_2021}, which are readily available commercially. In this review, we exclusively focus on solid-state systems with particular emphasis on the superconducting platform.

%\bt{\blue{Challenges of RT electronics; Motivation of Cryo ICs}}. 
After approximately three decades of development, solid-state quantum systems, particularly superconducting ones, have entered the Noisy Intermediate-Scale Quantum device stage. The fidelity of the physical qubit has now reached the threshold required for implementing surface-code-based quantum error correction~\cite{FowlerPRA2012}. However, constructing a valuable fault-tolerant system necessitates several million physical qubits~\cite{ChakrabortyJSSC2022}. This poses a challenge due to the limitations of state-of-the-art control and readout (C/R) room-temperature (RT) electronic equipment. The number of cables/transmission lines would be roughly twice the number of qubits, resulting in an unmanageable interconnection complexity and significant crosstalk. Additionally, connecting transmission lines between RT electronics on racks~\cite{Reilly_npjQI_2015} and quantum processors in cryostats introduces unnecessary latency and thermal noise. To overcome these limitations, researchers have proposed and extensively investigated the concept of cryogenic C/R integrated circuits (ICs) as described in Refs.~\onlinecite{Charbon_ISSCC_2017, Charbon_SSCM_2021, Sebastiano2017,MukhanovIEDM2019} over recent years.

%\bt{\blue{Conception and classification}}. 
The Cryogenic C/R ICs, as their name suggests, are integrated versions of C/R electronic equipment deployed inside the cryostat at temperatures ranging from millikelvin to around 5 K (known as cryogenic temperatures). For convenience, we refer to these chips as Cryo ICs. Furthermore, based on the control methods employed, these Cryo ICs can be classified into two groups: radio-frequency (RF) and rapid single flux quantum (RSFQ) ICs. The RF ICs have a similar architecture to RT equipment and are typically based on commercial CMOS technology; however, a few of them utilize non-silicon technologies such as GaAs~\cite{LeiAPMC2018} or SiGe heterojunction~\cite{PengISSCC2022}. To maintain consistency with the RF and quantum computing communities, all CMOS-based ICs including those based on non-silicon technology are referred to as Cryo-CMOS ICs in this review. Additionally, RSFQ ICs represent a novel control scheme that utilizes SFQ pulses generated by RSFQ logic circuits for controlling superconducting qubits~\cite{McDermottPRA2014,LikharevIEEE1991}. Unlike conventional RF ICs, microwaves are not incorporated in RSFQ-based IC designs.

%\bt{\blue{Paper Organization}}. 
In this review, we comprehensively discuss both types of cryogenic C/R ICs with a specific focus on the well-established Cryo-CMOS ones. Specifically, our discussion encompasses the control and readout principles, system architecture, and designs of cryogenic CMOS ICs, as well as the characterization and modeling of field effect transistors (FETs) at CT and the fundamental principles of RSFQ circuits. The detailed organization of this review is outlined as follows

\begin{enumerate}[1)]
\item In Section~\ref{section_2}, RT C/R electronics are introduced, including the basic concepts of qubits in Section~\ref{section_2_1}, the microwave-based control and readout principle in Section~\ref{section_2_2}, the system structure of RT C/R electronics in Section~\ref{section_2_3}, RF performance indices in Section~\ref{section_2_4}.
\item Section~\ref{section_3} provides a detail review of Cryo-CMOS ICs, including the proposed structure (Section~\ref{section_3_1}), Characterization and modeling of MOSEFT in cryogenic temperatures (Section~\ref{section_3_2}), the designs of Cryo-CMOS ICs and their performances (Section~\ref{section_3_3}). Specifically, in Section~\ref{section_3_2}, characterization results covering DC, RF electrical properties, and other cryogenic effects are shown in Section~\ref{section_3_2_1}, the physics-based, compact, and RF models of MOSFET are disscussed in Section~\ref{section_3_2_2}. In Section~\ref{section_3_3}, the designs of pulse modulators, RF receivers, LO generation module, and multiplexers are reviewed in the first four subsections, while the final subsection of Section~\ref{section_3_3} (Section~\ref{section_3_3_5}) reviews the performances of entire prototype chips that discussed in the first two subsections. 
\item Section~\ref{section_4} offers a brief introduction to the RSFQ technology, including basic concepts (Section~\ref{section_4_2}) and its qubit control principle (Section~\ref{section_4_2}). Furthermore, the potential to build RFSQ-based Cryo ICs is discussed in Section~\ref{section_4_2}.
\item The final section provides a summary and outlook of Cryogenic ICs.
\end{enumerate}

\section{Introduction to Room-Temperature Control and Readout Electronics}\label{section_2}

%\bt{\blue{Basic conceptions of control and readout electronics}}. 
A typical solid-state quantum system comprises two main components: the quantum processor, which serves as the carrier of qubits and stores/processes quantum information; and the classical part, encompassing the surrounding classical electronics such as cables and central computer. In this section, we focus on one specific aspect of the classical part: RT C/R electronics. This component facilitates translation of classical instructions from the central computer into microwaves that can be recognized by qubits or vice versa for readout microwaves from qubits back to the central computer. Some references also refer to this component as quantum-classical interference. For discussions on its cryogenic versions (Cryo-CMOS and RSFQ), please refer to Section~\ref{section_3} and Section~\ref{section_4}.

\begin{figure*}
\centering
\includegraphics[width=0.95\linewidth]{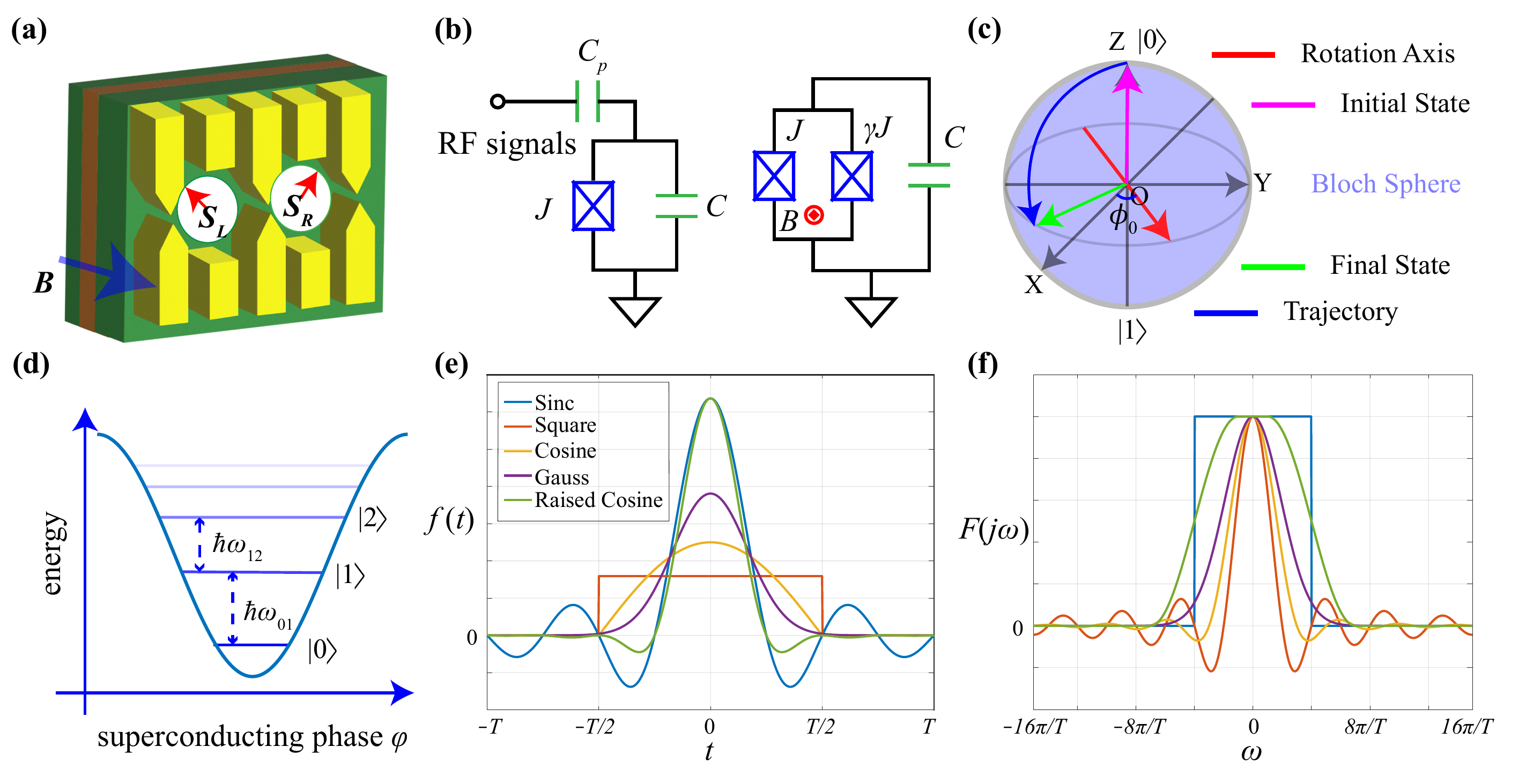}
\caption{\bt{Control principle and structure of qubits.} (a) Structure of Loss-Divincenzo spin qubit. (b) Structure of transmons, the one with fixed frequency on the left, the one with tunable frequency on the right. (c) Effect of microwave pulse on Bloch sphere (d). Level structure of transmon. (e) Time domain diagram of envelopes. (Their integral values over $\mathbb{R}$ are set to 1. The width of finite-length envelopes $T=3\sigma$ for Gauss envelope, where $\sigma$ is the standard deviation of Gauss envelope. The second zero point of the raised cosine envelope is set to $T/2$) (f) Frequency domain diagram of envelopes. Panel (a) is from Ref.~\cite{Burkard_Rev_Mod_Phys_2023}}.\label{fig1}
\end{figure*}

\subsection{Basic Concepts of Qubits}\label{section_2_1}

%\bt{\blue{General Statement}}. 
The basic concepts of qubits are briefly introduced in this subsection prior to the formal introduction of control and readout electronics. As mentioned in~\nameref{Intro}, two prominent solid-state qubits are considered: spin qubits in the semiconductor platform and transmons in the superconductor platform.

%\bt{\blue{Spin qubits}}. 
The main structure of spin qubits, quantum dots, are semiconductor nanostructures surrounded by a different kind of semiconductor with a distinct band gap. According to statistical physics, the entire structure of quantum dots shares the same Fermi energy level; however, the different band gap creates a potential well in the conduction band. Due to the quantum confinement effect, separated energy levels can exist in this potential well and trap excited electrons in the conduction band. The spin freedom of these trapped electrons is used to encode quantum information and serves as a natural qubit. As an example, Fig.~\ref{fig1}(a) depicts the structure of a Loss-Divincenzo single spin qubit~\cite{LossPRA1998}.

%\bt{\blue{Transmons}}. 
Another type of qubit is the transmon, composed of superconducting Josephson junctions. The term "transmon" refers to transmission line shunted plasma oscillation qubits, and its structure is illustrated in Fig.~\ref{fig1}(b). In this configuration, a Josephson junction or a DC superconducting quantum interference device (DC SQUID) is coupled to a large capacitor for phase noise suppression. The effective Hamiltonian of the structure with a DC SQUID can be expressed as follows
\begin{equation}
\hat{H} = 4E_{\rm C} n^2 -E_{\rm J\Sigma}\sqrt{\cos^2(\phi_{\rm e})+d^2\sin^2(\phi_{\rm e})}\cos(\phi),
\end{equation}

where $E_C$ represents the capacitive energy, $E_{J\Sigma}$ is the sum of the energies of the two Josephson junctions, $\phi_e = \pi\Phi_{\rm ext}/\Phi_0$ denotes the reduced external magnetic flux with the real external magnetic flux $\Phi_{\rm ext}$, and the flux quanta $\Phi_0=h/2e$. The potential of this Hamiltonian, as illustrated in Fig.~\ref{fig1}(d), gives rise to an energy level structure characterized by distinct energy gaps between adjacent levels. The computational subspace can be defined as the span of the states corresponding to the lowest two energy levels, denoted as $\bf{span}\left\{|0\rangle, |1\rangle\right\}$. Notably, the transition frequency between these lowest two levels falls within the microwave frequency range (typically around 5 GHz for transmons). Additionally, it is worth mentioning that the control frequency of spin qubits can also be interpreted as the transition frequency between two specific levels, typically operating at around 10 GHz.

\subsection{Microwave-based Control and Readout Principle }\label{section_2_2}

%\bt{\blue{Importance of the microwaves}}. 
Microwaves play a pivotal role in manipulating quantum states in solid-state materials, making them indispensable for cutting-edge RT C/R electronics and Cryo-CMOS ICs that rely on the microwave-based C/R principle. In this subsection, we present an introduction to the microwave-based C/R principle along with commonly employed waveforms of control signals.

%\bt{\blue{Control signals}}. 
The control of qubits encompasses the preparation of initial states and the implementation of one-bit or two-bit gates. Fortunately, nearly all control processes in both transmons and spin qubits can be achieved solely using microwaves, except for the external magnetic flux required for tunable transmons/couplers (which should be regulated by DC currents). Additionally, recent research has demonstrated an all-microwave controlled scheme~\cite{ShiraiPRL2023}. To elucidate the utilization of microwaves in qubit control, let us examine the realization of Pauli X and Y gates in transmons. In most scenarios, a single transmon qubit is capacitively coupled to transmission lines (Fig.~\ref{fig1}(b)). The microwave signal is delivered through the transmission line with the following waveform
\begin{equation}
f(t) = s(t)\cos(\omega t +\phi_0),
\end{equation}
where $s(t)$ represents the envelope of signal with limited length, $\omega$ is the frequency of local oscillator (LO), and $\phi_0$ is the initial phase of LO. By considering the interactive Hamiltonian between microwaves and qubits, we can express the unitary evolution operator as
\begin{equation}
\begin{aligned}
&\hat{U}^{\phi_0}_{\rm RF,d}(t)\\
&=\exp\left(\left[\frac{i}{2}\Omega \int_0^{\rm t}s(t^\prime)dt^\prime\right]\cos(\phi_0)\sigma_{\rm x}+\sin(\phi_0)\sigma_{\rm y}\right)
\end{aligned},
\end{equation}
where The Rabi oscillation frequency, denoted as $\Omega$, is determined by the coupling strength between the transmission line and transmons. As illustrated in Fig.~\ref{fig1}(c), this operator induces a clockwise rotation of the state on the Bloch sphere along the axis vector $(\cos(\phi_0),\sin(\phi_0),0)$ by an angle of $\Omega \int_0^t s(t^\prime) dt^\prime$ radians. By setting $\phi_0=0$ and rotating the angle to $\pi$, we can achieve implementation of the Pauli X gate. Similarly, utilizing the same approach allows for realization of the Pauli Y gate.

\begin{table*}[width=0.7\textheight,pos=!h]
\caption{Time/Frequency domain expressions of four envelopes ($\alpha$,$\beta$, $\gamma$, $a$ and $T$ are variable coefficients, $sinc(x)\equiv \sin(x)/x$)}
\label{table1}
\begin{tabular*}{\tblwidth}{@{} LLL@{} }
\toprule
Envelopes & Time Domain & Frequency Domain\\
\midrule
\makecell[l]{Rectangle} & $f(t) =(1/\gamma)rect(t/\gamma)$ & $F(j\omega) =sinc(\gamma\omega/2)$ \\ \midrule
\makecell[l]{Cosine} & $f(t) =(1/2)\cos(\beta t)$ & $F(j\omega)=(1/4\pi)\left\{sinc[(\pi/2\beta)(\omega-\beta)]+sinc[(\pi/2\beta)(\omega+\beta)]\right\}$ \\ \midrule
\makecell[l]{Gauss} & $f(t) =\sqrt{\alpha/\pi}\exp(-\alpha t^2)$ & $F(j\omega) = \exp(-\omega^2/4\alpha)$ \\ \midrule
\makecell[l]{Raised \\ Cosine} & \makecell[l]{$f(t) = (1/T)sinc(\pi t/T)\cdot$\\ $\cos{(a\pi t/T)}/[1-(2a t/T)^2]$} & \makecell[l]{$F(j\omega)=\begin{cases}1&|\omega|\leq\pi(1-a)/T\\ \cos(\pi T/2a)|\omega|/(2\pi-(1-a)/2T)^2 &\pi(1-a)/T<|\omega|<\pi(1+a)/T\\ 0 &else\end{cases}$}\\
\bottomrule
\end{tabular*}
\end{table*}

%\bt{\blue{Readout Principle}}. 
Next, we delve into the concept of dispersive readout and its connection to microwaves. Dispersive readout is based on cavity electrodynamics~\cite{Scully_zubairy_1997} (cavity QED) and its derivative circuit QEC~\cite{Blais_Rev_Mod_Phys_2021}. In this approach, a qubit is coupled with a microwave cavity that operates at a frequency far-off from the qubit's resonance, such as a half-wavelength coplanar waveguide or an LC circuit. By injecting a resonant microwave pulse into the cavity, two coherent states with distinct dressed frequencies are generated within it. This process can be comprehended through the far-off-resonance Hamiltonian
\begin{equation}
H_{\rm disp}\simeq (\omega_r+\chi\sigma_z)(a^\dagger a+\frac12)+\frac{\tilde{\omega}_q}{2}\sigma_z,
\end{equation}
where the $\omega_r$ and $\omega_q$ stands for frequencies of resonator and qubit correspondly, $\chi$ is the qubit-state dependent frequency shift. Based on this Hamiltonian, it can be inferred that the dressed states are determined independently by the qubit states $|0\rangle$ and $|1\rangle$, with frequencies $\omega_r-\chi$ and $\omega_r +\chi$ respectively. Once the dressed state is formed, a weaker detection signal is injected, causing the reflected microwaves to carry information about the qubit in terms of amplitude and phase following the collapse of the quantum state triggered by pulse injection. Ideally, for a single qubit, the reflected microwave $S_{21}(t)$ represents an amplitude-phase-modulated single frequency signal given as $S_{21}(t) = A_{|0\rangle/|1\rangle}\cos(\omega_{\rm r} t +\phi_{|0\rangle/|1\rangle})$, where indices of $A$ and $\phi$ indicate their state-dependent properties, and $\omega_{\rm r}$ denotes the readout signal frequency.

%\bt{\blue{Different envelopes and DRAG signals}}. 
Finally, we discuss the waveforms of control signals and their reduction by the adiabatic gate (DRAG). In our discussion, waveforms represent the envelopes of control signals $s(t)$ with the aim of suppressing leakage outside of the computation subspace, specifically leakage out of the $|0\rangle$ and $|1\rangle$ states. For transmons, it is typical for the transition frequency between the $|1\rangle$ and $|2\rangle$ states ($\omega_{12}$) to be 100-300 MHz smaller than the transition frequency within the computation subspace. Therefore, in order to achieve a certain fidelity level, it is necessary to minimize suppression on $\omega_{12}$. Fig.~\ref{fig1}(e) shows five typical waveform envelopes (with corresponding analytic expressions provided in Tab.~\ref{table1}), where their widths and integral values over $\mathbb{R}$ (representing rotation angle) are set equal. The Fourier transformations are shown in Fig.~\ref{fig1}(f). Theoretically, a sinc function envelope ($s(t)=\beta\sin(\alpha t)/(\alpha t)$) has zero leakage; however, its implementation in electronics is challenging so Gauss or raised cosine envelopes are commonly adopted instead~\cite{Razavi2011}. Furthermore, to further enhance leakage suppression efficiency, Refs.~\onlinecite{MotzoiPRL2009, GambettaPRA2011} proposed an approach called derivation reduced by adiabatic gate (DRAG), which involves using a microwave control signal with a specific form
\begin{equation}
f(t) = s(t)\cos(\omega t+ \phi_0) + \lambda \frac{\dot{s}(t)}{\alpha}\sin(\omega t + \phi_0),
\end{equation}
where $s(t)$ represents the main envelope as defined in Equation (2). $\dot{s}(t)$ denotes the derivative of $s(t)$ with respect to time. $\alpha$ and $\lambda$ are variable coefficients, and their optimal values can be determined through experiments.

\subsection{System Structure of Room-temperature Control and Readout Electronics}\label{section_2_3}

%\bt{\blue{General statement}}. 
The transceiver serves as a fundamental component in RF circuits and communication equipment, exhibiting a highly consistent system structure with both the RT C/R electronics and Cryo-CMOS ICs. In this subsection, we delve into the system structure of RT C/R electronics from the perspective of the transceiver while emphasizing their distinguishing features.

\begin{figure*}
\centering
\includegraphics[width=0.95\linewidth]{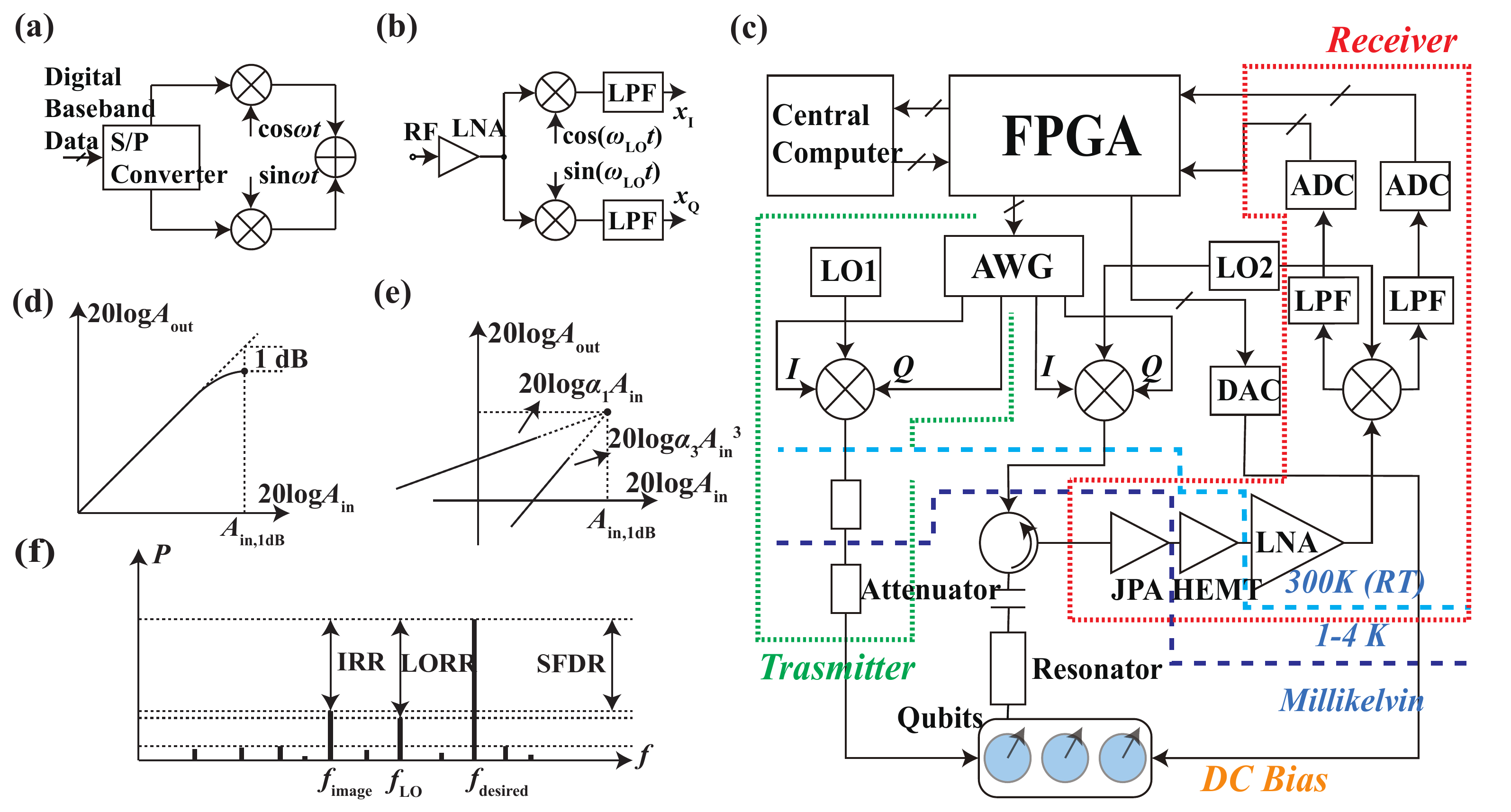}
\caption{\bt{RT C/R electronics.} (a) Direct conversion transmitter for QPSK. (b) Sliding intermediate frequency receiver. (c) System structure of RT C/R electronics. (d) 1 dB gain compression point. (e) 3rd order intermodulation point. (f) IRR, LORR, and SFDR.}\label{fig2}
\end{figure*}

%\bt{\blue{Transmitter/Control signal genration}}. 
A transceiver comprises a transmitter and a receiver. In modern digital communication technology, both components are based on quadrature modulation/demodulation techniques~\cite{Rodger1985}. Specifically, two commonly employed schemes include Quadrature Amplitude Modulation and Quadrature Phase Shift Keying~\cite{Aiysha2019}. In the latter scheme, the system architecture of a transmitter is depicted in Fig.~\ref{fig2}(a). Initially, digital signals are converted into two baseband signals denoted as $I(t)$ and $Q(t)$ through a serial-parallel converter (S/P converter). These signals are then individually multiplied with a local oscillator (LO) comprising of a cosine component $\cos(\omega_{\rm LO} t)$ and a quadrature component $\sin(\omega_{\rm LO} t)$. Finally, the two multiplied signals are combined to obtain the resultant modulated microwave signal, which reads
\begin{equation}
x(t) = I(t)\cos(\omega_{\rm LO} t) - Q(t)\sin(\omega_{\rm LO} t).
\end{equation}
The signal bears resemblance to the control signal as depicted in Equation (2). Specifically, $I(t)$ corresponds to $s(t)\cos(\phi_0)$, while $Q(t)$ corresponds to $s(t)\sin(\phi_0)$. In the QPSK modulation scheme, both baseband signals, namely $I(t)$ and $Q(t)$, manifest as square wave signals that alternate between binary positive and negative levels. However, in order to attain the envelope signal denoted by $s(t)$ in Equation (2), it is imperative to substitute the S/P converter with two digital-to-analog converters (DAC) connected to a read-only memory (ROM) housing the envelope waveform.

%\bt{\blue{Reciever/Readout}}. 
The receiver performs the inverse function of the transmitter, known as down-conversion in communication technology. Fig.~\ref{fig2}(b) illustrates a sliding intermediate frequency (IF) receiver~\cite{Razavi2011}, which belongs to the category of heterodyne receivers. In this configuration, the local oscillator (LO) frequency can be adjusted to achieve baseband conversion when $f_{\rm LO1}+ 1/2 f_{\rm LO1}=f_{\rm in}$. For quantum applications, single down-conversion with non-zero intermediate frequencies is adequate since subsequent analog-to-digital converters (ADCs) and digital signal processing units can handle the resulting low-frequency signals.

%\bt{\blue{Summary of RT C/R electronics}}. 
To depict the comprehensive system architecture of RT electronics for superconducting transmons, Fig.~\ref{fig2}(c) integrates the transmitter (control devices) and the receiver (readout devices). Furthermore, the RT electronics encompass Josephson Parameter Amplifiers (JPA), attenuators, and circulators in addition to the transceiver.

\subsection{Nonideal Effects in RF electronics}\label{section_2_4}
%\bt{\blue{General statement}}. 
In RF electronics, nonlinearity and other phenomena can lead to undesired noise generation and loss of amplification. To comprehensively comprehend these nonideal effects, it is crucial to investigate their underlying causes and associated performance indicators. Within this subsection, we will present five nonideal effects along with their corresponding performance indicators: harmonic distortion, gain compression, intermodulation, LO leakage, and image issues.

% \bt{\blue{Harmonic distortion and Gain compression}}. 
The impact of system nonlinearity on a single input signal, denoted as $x(t)$, can be described by harmonic distortion and gain compression. Assuming the system is memoryless, the output signal $y(t)$ can be approximated as $y(t)\approx\alpha_1 x(t)+\alpha_2 x^2(t) + \alpha_3 x^3(t)$~\cite{Razavi2011}. When the input signal $x(t)$ is a single tone ($x(t)=A\cos(\omega t)$), the output signal can be expressed as follows
\begin{equation}\begin{aligned}
y(t)&=\frac{\alpha_2 A^2}{2}+(\alpha_1A+\frac{3\alpha_3A^3}{4})\cos(\omega t)\\
&+\frac{\alpha_2A^2}{2}\cos(2\omega t)+\frac{\alpha_3 A^3}{4}\cos(3\omega t)
\end{aligned}.
\end{equation}
The additional terms $\cos(2\omega t)$ and $\cos(3\omega t)$ represent harmonic distortion, while the constant term is referred to as DC offset. Focusing on the term $cos(\omega t)$, if the coefficients $\alpha_1$ and $\alpha_3$ satisfy $\alpha_1\alpha_3<0$, the gain $(\alpha_1 A+\frac{3\alpha_3 A^3}{4})$ will decrease as the input amplitude $A$ increases. This phenomenon can be quantified by measuring the 1 dB gain compression point. In Fig.~\ref{fig2}(d), this point corresponds to an input amplitude $A_{in,1 dB}$ where the actual output gain decreases by 1 dB compared to the ideal output gain. An approximate expression for $A_{in,1 dB}$ is given by
\begin{equation}
A_{\rm in,1 dB}=\sqrt{0.145\left|\frac{\alpha_1}{\alpha_3}\right|}.
\end{equation}

%\bt{\blue{Intermodulation}}. 
Intermodulation is a nonideal effect that arises from the interaction between two strong interfering signals within a communication system. To characterize this phenomenon, a two-tone signal of the form $x(t)=A_1\cos(\omega_1 t)+A_2\cos(\omega_2 t)$ is applied at the input of the system. By considering the input-output relationship of the previously mentioned nonlinear system, four third-order intermodulation terms can be derived (for detailed information, refer to~\onlinecite{Razavi2011}). The third-order intercept point (IP3) is commonly employed to describe this effect. In Fig.~\ref{fig2}(e), when both tones have equal input amplitudes ($A_1=A_2$), the intersection point between the intermodulation term and signal term represents IP3. Similar to the 1 dB compression point, we denote the input amplitude at IP3 as $A_{IP3}$ and it can be approximated as
\begin{equation}
A_{\rm IP3}=\sqrt{\frac{4}{3} \left|\frac{\alpha_1}{\alpha_3}\right|}.
\end{equation}

\begin{figure}
\centering
\includegraphics[width=0.95\linewidth]{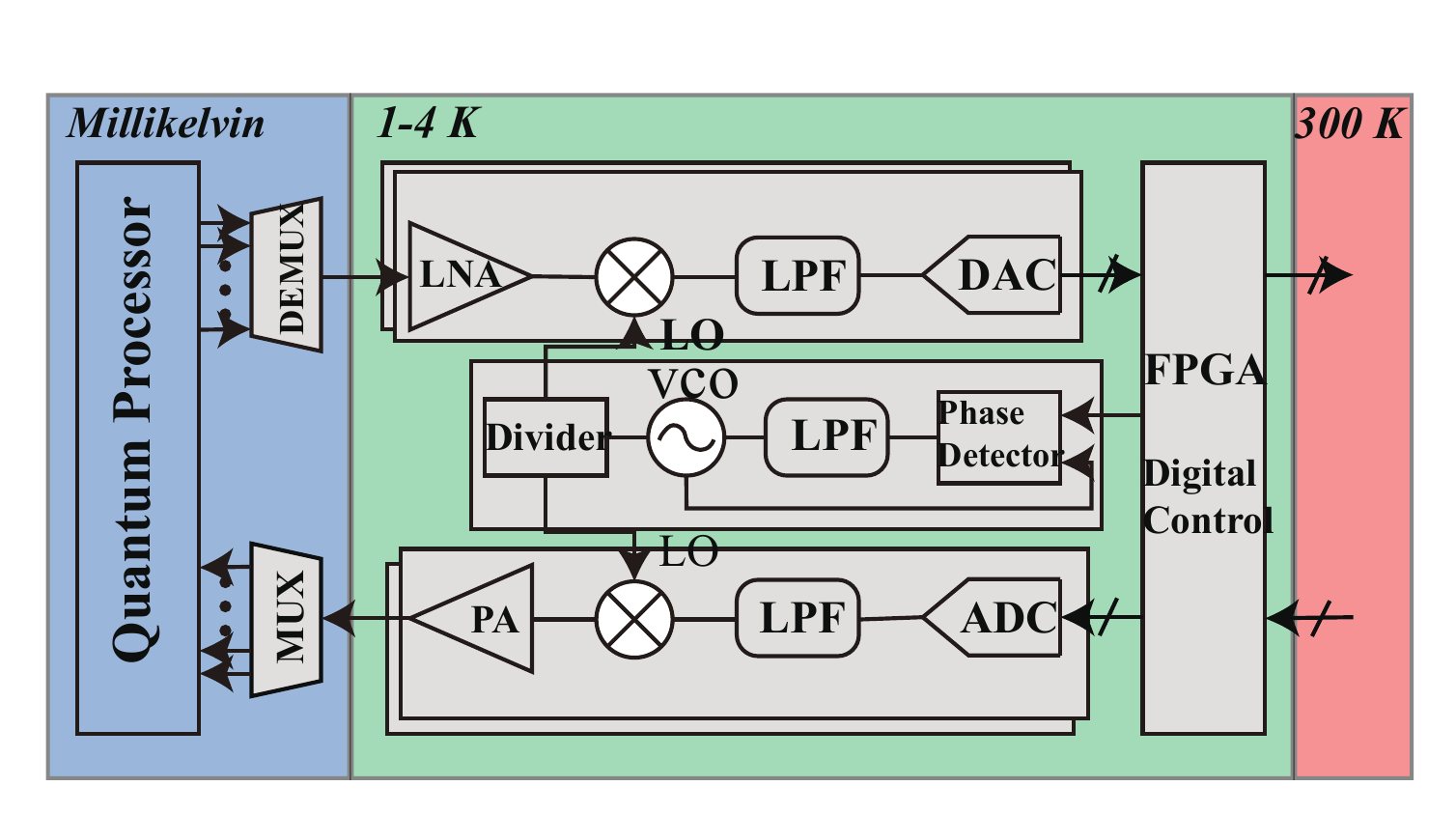}
\caption{\bt{Proposed System Structure of Cryo-CMOS ICs.} Figure is from Ref.~\cite{Charbon_ISSCC_2017}.}\label{fig2n}
\end{figure}

%\bt{\blue{LO leakage}}. 
LO leakage refers to the ingress of LO signals into the input port or antenna in a direct conversion transmitter, resulting from substrate couplings or other channels. The LO rejection ratio (LORR) is employed as a metric to quantify this phenomenon. In Fig.~\ref{fig2}(f), assuming a single-tone signal is injected into the system, the attenuation of the output LO component relative to the desired signal in the output power spectrum serves as an indicator for LORR.

%\bt{\blue{Image problem}}. 
The image issue arises as a result of the mixer. Assuming the LO frequency of the receiver is $\omega_{\rm LO}$ and the frequency of the output signal is $\omega_{\rm in}$, where $\omega_{\rm in}>\omega_{\rm LO}$. The image frequency $\omega_{\rm im}=2\omega_{\rm LO}-\omega_{\rm in}$ is the reflection of $\omega_{\rm in}$ around $\omega_{\rm LO}$. In the power spectrum, interference at the frequency of $\omega_{\rm im}$ is also down-converted to the desired frequency (as the power spectrum is a single-side spectrum). The image rejection ratio (IRR) is used to measure this effect. In Fig.~\ref{fig2}(f), the attenuation of the image component compared to the desired signal represents the IRR. 

%\bt{\blue{SFDR}}. 
Finally, a more comprehensive metric is introduced: the spurious free dynamic range (SFDR). In Fig.~\ref{fig2}(f), when the input signal remains a single-tone signal, the attenuation of the most detrimental spurious noise relative to the desired signal within the designated bandwidth represents the SFDR.

\section{Cryogenic CMOS ICs}\label{section_3}

This section provides a comprehensive discussion on Cryo-CMOS ICs, encompassing the proposed structure of Cryo-CMOS ICs, characterization and modeling of MOSFET at cryogenic temperatures, as well as the designs and performances of recent Cryo-CMOS prototype chips.

\subsection{General Structure}\label{section_3_1}

%\bt{\blue{General statement}}. 
As previously mentioned in \nameref{Intro}, the concept of Cryo-CMOS ICs was initially proposed by E. Charbon from Delft University of Technology~\cite{Charbon_ISSCC_2017, Charbon_SSCM_2021, Sebastiano2017}. In this context, we will discuss the suggested general structure along with its associated benefits and challenges as a foundation for further discourse.

%\bt{\blue{Depiction of System structure}}. 
The proposed system architecture for Cryo-CMOS ICs is depicted in Fig.~\ref{fig2n}, resembling that of RT electronics, comprising transmitters (control) and receivers (readout), accompanied by optional DC bias generation blocks catering to various qubit types. Furthermore, a digital component encompassing FPGA, microprocessors, and memories has been incorporated to facilitate expedited waveform loading and feedback control for fault-tolerant designs.

%\bt{\blue{Benefits}}. 
The entire C/R electronics is housed within the cryostat in Fig.~\ref{fig2n}, offering several advantages, including enhanced quantum gate fidelity and reduced thermal noise of RF electronics. Besides, despite being approximately one meter away from the quantum processor, the 4 K stage of the refrigerator effectively minimizes latency compared to the RT electronics. Given sufficient refrigeration capacity, it becomes feasible to incorporate complex digital components into either the 4 K stage or even the millikelvin stage, resulting in negligible latency. Furthermore, commercial CMOS technology can naturally address scalability issues due to its high level of integration.

%\bt{\blue{Challenges}}. 
Besides the benefits, cryogenic temperatures also present challenges that can complicate designs. The first challenge arises from the lack of reliable commercial Metal-Oxide-Semiconductor Field-Effect Transistor (MOSFET) models for IC designs, as the current foundry-provided model fails to simulate MOSFETs at temperatures of a few Kelvins. Another challenge stems from higher threshold voltages and increased flicker noise at cryogenic temperatures~\onlinecite{MehrpooISCAS2019}, which may pose difficulties in designing certain blocks such as Low Noise Amplifiers (LNA), Oscillators, and Phase-Locked Loops (PLL). Section~\ref{section_3_2} will discuss the extensive recent studies addressing the first challenge, while Section~\ref{section_3_3} will address the second challenge.

\subsection{MOSFET in Cryogenic Temperature}\label{section_3_2}
\begin{figure*}
\centering
\includegraphics[width=1\linewidth]{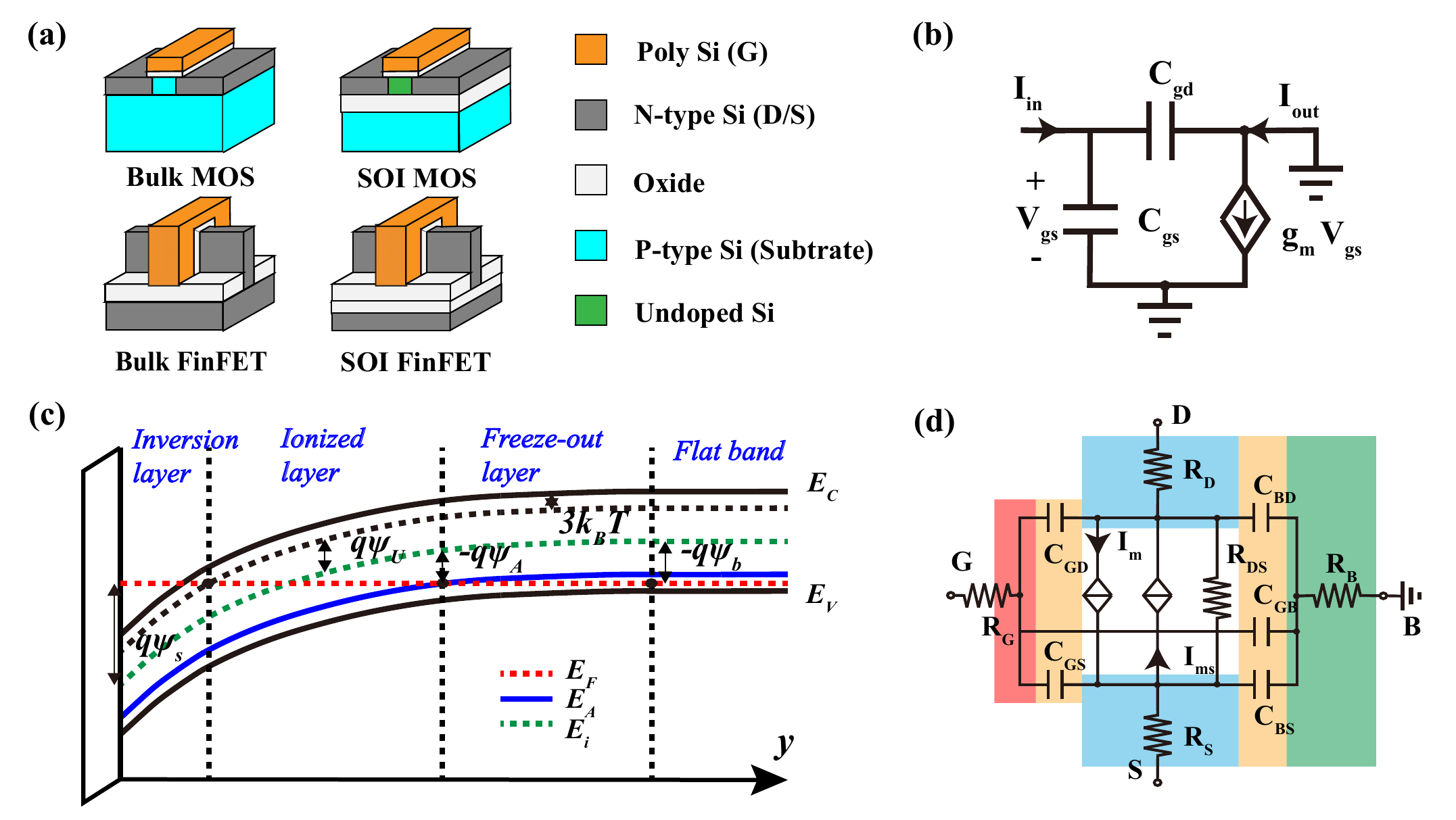}
\caption{\bt{Structure, models, and band structure of MOSFET.} (a) Geometric structures of FETs. (b) Simplest small signal model. (c) Band structure of MOSFET at cryogenic temperature. (Where the y-axis represents the depth into the substrate) (d) Small signal model of FDSOI MOSFET. Panel (c) is from Ref.~\cite{BeckersTED2018}, panel (d) is from Ref.~\cite{HanESSDERC2022}.}\label{fig3}
\end{figure*}

%\bt{\blue{General Statement}}. 
This subsection presents an analysis of MOSFET performance at cryogenic temperatures. Prior to delving into the subject, it is essential to introduce the concept of semiconductor technology nodes. Despite slight variations among different foundries, FETs with similar process nodes exhibit a predominantly consistent geometric structure. In this review, particular attention should be directed towards four distinct FET structures: bulk MOS, SOI MOS, bulk FinFET, and SOI FET. The schematic representations of these four FET structures are depicted in Fig.~\ref{fig3}(a).

\subsubsection{Characterization Results}\label{section_3_2_1}

%\bt{\blue{General statement}}. 
This subsection presents a comprehensive overview of previous findings pertaining to the characterization of FETs, with a specific focus on various aspects of DC electrical parameters such as transconductance ($g_{\rm m}$), threshold voltages ($V_{\rm T}$), subthreshold slope ($SS$), and mismatch. Furthermore, it delves into RF electrical parameters like the unit current gain cutoff frequency ($f_{\rm T}$) and investigates the influence of cryogenic effects on FET performance.

\paragraph*{\textbf{DC Electrical Properties}}

%\bt{\blue{Transconductance and Threshold}}. 
Firstly, we present the fundamental DC electrical characteristics of FETs: transconductance ($g_{\rm m}$) and threshold voltage ($V_{\rm T}$). These parameters are extensively discussed across various technology nodes. At cryogenic temperatures, FETs exhibit enhanced transconductance in the saturation region and elevated threshold voltage due to reduced electron scattering and increased ionization energy. Table~\ref{table2} provides the characterization results pertaining to these properties.

%\bt{\blue{Subthreshold slope}}. 
The subthreshold slope ($SS$) is another crucial parameter that characterizes the switching speed of FETs. It quantifies the change in drain current ($I_{\rm DS}$) for a tenfold variation in gate voltage ($V_{\rm G}$). Mathematically, it can be described as follows
\begin{equation}
SS = \frac{dV_{\rm G}}{d\log(I_{\rm DS})}.
\end{equation}
Typically, both digital and analog designs aim for lower values of $SS$. Fortunately, at cryogenic temperatures, $SS$ is generally lower compared to room temperature. However, this finding contradicts the theoretical prediction of $SS=\ln(10)n k_{\rm B} T/q$ (where $k_{\rm B}$ represents the Boltzmann constant and $q$ denotes the charge of an electron). For example, in Ref.~\onlinecite{IncandelaJOEDS2018}, the improvement in $SS$ from 300K to 4K is only 3.8 times, significantly below the theoretical prediction of over 100 times. This inconsistency may be attributed to the temperature dependence of the slope factor ($n$), which has been determined based on experimental results mentioned in Ref.~\onlinecite{IncandelaJOEDS2018}. Physics-based explanations have also been proposed in Refs.~\onlinecite{BeckersTED2018, Beckers2022}, which will be discussed further in Section~\ref{section_3_2_2}.

%\bt{\blue{Introduction to Mismatch}}. 
Finally, the issue of mismatch is addressed. As a prevalent concern in IC design, mismatch arises due to non-idealities during micro/nano processing, with its impact worsening for nano-scale devices. Extensive research has been conducted on characterizing FET mismatch at room temperature (RT)~\cite{Marcel1991,Jeroen2005}, while cryogenic investigations have also been performed on 40-nm bulk CMOS FETs~\cite{HartESSDERC2018,HartJEDS2020}. In the subsequent discussion, we focus specifically on the cryogenic findings.

%\bt{\blue{Results of Mismatch}}. 
The three parameters discussed in Ref.~\onlinecite{HartESSDERC2018} are $\sigma(\Delta V_{\rm T})$, $\sigma(\Delta I_{\rm D}/I_{\rm D})$, and $\sigma(\Delta \beta/ \beta)$ (where $\beta$ represents the transconductance factor). Additionally, Ref.~\onlinecite{HartJEDS2020} also has discussed the $\sigma(\Delta SS/SS)$. The findings from Refs.~\onlinecite{HartESSDERC2018,HartJEDS2020} can be summarized as follows: 1. Similar to room temperature FETs, at cryogenic temperatures, mismatch in the weak/moderate inversion region is more pronounced than in the strong inversion region. 2. At 4K, $\sigma(\Delta I_{\rm D}/I_{\rm D})$ increases by approximately 1.5 times in the strong inversion region and by 10 times in the weak/moderate inversion region compared to room temperature conditions. 3. When FETs operate at a constant $g_{\rm m}/I_{\rm D}$ biasing point, mismatch only increases by a factor of 1.2 times. These results confirm the presence of significant mismatch at cryogenic temperatures; however, it is fortunate that this increase is only observed when biased at constant $g_{m}/I_D$ (which corresponds to most analog circuits). In addition to characterizing these results, effective models for quantifying mismatch are also discussed in Refs.\onlinecite{HartESSDERC2018,HartJEDS2020}, which will be concisely addressed in the next subsection.

\paragraph*{\textbf{RF Electrical Properties}}

%\bt{\blue{General statement}}. 
Compared to the well-studied DC electrical properties, RF characteristics have received relatively less attention due to the intricacies involved in their measurement, particularly S-parameter measurement. However, there are relevant studies available such as those mentioned in Refs.~\onlinecite{ChakrabortyIEEE2021, TangJEDS2022,HanESSDERC2022}, which explore different semiconductor processing techniques. Specifically, references~\cite{ChakrabortyIEEE2021, HanESSDERC2022} employ 22-nm FDSOI CMOS processing while reference~\cite{TangJEDS2022} utilizes 40-nm bulk CMOS processing.

%\bt{\blue{Methods of chracterlization (S parameters)}}. 
The RF properties of FETs are characterized through the analysis of S, Z, or Y parameters. For n-type Metal-Oxide-Semiconductor FETs (MOSFETs), the transistor is biased to a specific region, either saturation or linear region, and small high-frequency signals are applied to the gate and drain terminals. Since the source and substrate are grounded, the voltages at the two ports correspond to $V_{\rm GS}$ and $V_{\rm DS}$. This same port configuration is employed for p-type MOSFETs as well. Consequently, MOSFETs are treated as two-port devices for RF characterization purposes. Vector network analyzers are commonly utilized in practice for measuring S parameters. The findings reported in Refs.~\onlinecite{ChakrabortyIEEE2021, TangJEDS2022,HanESSDERC2022} demonstrate that both $S_{11}$ and $S_{22}$ curves exhibit minimal sensitivity towards temperature fluctuations, thereby resulting in temperature-invariant impedance.

\begin{table*}[width=0.7\textheight,pos=!h]
\caption{Cryogenic DC properties of FETs (compared with RT)}
\label{table2}
\begin{tabular*}{\tblwidth}{@{} LLLLL @{} }
\toprule
~ & \cite{IncandelaJOEDS2018} & \cite{HartESSDERC2018} & \cite{ChakrabortyIEEE2021} & \cite{BeckersJOEDS2018} \\
\midrule
Temperature & 4 K & 4.2 K & 5.5 K & 4.2 K \\ \midrule
\makecell[l]{Processing \\ Node} & \makecell[l]{40-nm \\bulk CMOS} & \makecell[l]{40-nm \\bulk CMOS} & \makecell[l]{22-nm FDSOI \\CMOS} & \makecell[l]{28-nm \\ bulk CMOS} \\ \midrule
\makecell[l]{Transconductance $g_{\rm m}$ \\ /Mobility $\mu$} & $\times$2 & ~ & \makecell[l]{nFET: 133\% \\ pFET: 125\%} & nFET $\times$1.3 \\ \midrule
\makecell[l]{Threshold \\ voltage $V_{\rm TH}$} & 130\% & \makecell[l]{nFET: 100 mV \\ pFET: 180 mV} & \makecell[l]{nFET: 122.5 mV higher \\ pFET: 160 mV higher} & 0.1-0.2 V \\ \midrule
\makecell[l]{Subthreshold \\ slope $SS$} & 1/3.8 & ~ & ~ & \makecell[l]{11 mV/dec \\ (85\% smaller)} \\
\toprule
\cite{HanESSDERC2022} & \cite{LopezLAEDC2020} & \cite{SinghEDL2022} & \cite{ChabaneESSCIRC2021} & \cite{HanESSCIRC2021}\\ \midrule
3.3 K & 340 mK & 3.4 K & 4.6 K & 2.95 K \\ \midrule
\makecell[l]{22-nm \\ FDSOI CMOS} & \makecell[l]{14-nm \\SOI FinFET} & \makecell[l]{65-nm \\Bulk CMOS} & \makecell[l]{14-nm \\bulk FinFET} & \makecell[l]{16-nm \\bulk FinFET} \\ \midrule
~ & ~ & ~ & 133\% & \makecell[l]{$\times$2.2 (L=240 nm) \\ $\times$ 1.5 (L=16 nm)} \\ \midrule
70 mV & 143\% & \makecell[l]{nFET: 0.45 V\\ pFET: 0.2 V} & 80 mV & 12 mV \\ \midrule
\makecell[l]{nFET 22 mV/dec \\ pFET 33 mV/dec} & 22 mV/dec & ~ & 20 mV/dec & 18 mV/dec \\
\bottomrule
\end{tabular*}
\end{table*}

%\bt{\blue{Concepts of some important FoMs}}. 
In this section, we discuss the figures of merit (FoMs) for the unity current-gain cutoff frequency ($f_{\rm T}$) and the maximum oscillation frequency ($f_{\rm MAX}$). These FoMs represent the frequencies at which both the current gain and power gain are equal to one. However, direct measurement of these frequencies is not feasible due to their range being in a few hundred GHz, which surpasses the capabilities of available measurement instruments. Two feasible solutions exist: 1. Extrapolating limited measurement results; 2. Predicting frequencies through small signal models that consider parasitic interconnection, intrinsic and extrinsic parameters \cite{ChakrabortyIEEE2021,TangJEDS2022}. The second approach is more intricate as it requires consideration of various factors such as parasitic interconnections and intrinsic/extrinsic parameters \cite{HanESSDERC2022}. In Refs.~\onlinecite{ChakrabortyIEEE2021,TangJEDS2022}, de-embedding techniques like Short-Open-Load-Through method and Open-Short technique \cite{Koolen1991,EnzJSSC2000} have been employed to extract parameters in small signal models. However, Ref.~\onlinecite{HanESSDERC2022} neglects parasitic interconnections altogether. Referring to Ref.~\onlinecite{ChakrabortyIEEE2021}, we adopt the first method where short-circuit current gain ($H_{21}$) and Mason's unilateral power gain ($U$) are calculated using de-embedded S parameters, followed by inferring $f_{\rm T}$ and $f_{\rm MAX}$ through -20 dB/dec extrapolation from 1 GHz to 20 GHz results.

%\bt{\blue{Theoretical expression of FoMs}}. 
The simplified small signal model of a MOSFET is illustrated in Fig.~\ref{fig3}(b). In this study, we derive an analytical expression for the transit frequency ($f_{\rm T}$) and maximum frequency ($f_{\rm MAX}$) based on this model. By applying Kirchhoff's law, the input current $I_{\rm in}=j\omega (C_{\rm gs}+C_{\rm gd})V_{\rm gs}(\omega)$ and the output current $I_{\rm out} = (g_{\rm m}-j\omega C_{gd})V_{gs}(\omega)$. Consequently, the current gain can be expressed as $\left|I_{{out}}/I_{{in}} \right|= \sqrt{g_m^2+\omega^2C_{{gs}}^2}/( \omega(C_{{gs}}+C_{{gd}}))$, which approximates to $g_m/( \omega(C_{{gs}}+C_{{gd}}))$. Setting the current gain to unity yields an expression for $f_T$ as follows
\begin{equation}
f_{\rm T} = \frac{g_{\rm m}}{2\pi(C_{\rm gs}+C_{\rm gd})}.
\end{equation}
Another FoM $f_{\rm MAX}$ is closely connected with $f_{\rm T}$\cite{Physics_of_Semi_2006}, with expression as follow
\begin{equation}
f_{\rm MAX}=\sqrt{\frac{f_{\rm T}}{8\pi r_{\rm g} C_{\rm gd}}}.
\end{equation}

%\bt{\blue{Results and Analysis}}. 
According to the characterization results of $f_{\rm T}$ and $f_{\rm MAX}$, a significant enhancement of approximately 30\% is observed in both $f_{\rm T}$ and $f_{\rm MAX}$ when devices are operated at cryogenic temperatures. This improvement is in addition to the previously reported 70\% increase in p-MOS as mentioned in Ref.~\onlinecite{HanESSDERC2022}. These findings align with the analytical expression presented earlier, which highlights the temperature insensitivity of $C_{\rm gd}$ and $C_{\rm gs}$, along with an approximate 30\% rise in transconductance ($g_m$) as discussed during DC characterization. Another notable distinction is that both figures of merit (FoMs) for FDSOI MOSFETs~\cite{ChakrabortyIEEE2021,HanESSDERC2022} surpass those based on conventional bulk CMOS~\cite{TangJEDS2022}, indicating superior high-frequency performance achievable through SOI processing. Regarding cryogenic considerations, these enhanced FoMs would be advantageous for designing cryogenic ICs; however, it should be noted that modeling MOSFET behavior under such conditions may become more complex due to uncertain factors.

\paragraph*{\textbf{Cryogenic Effects}}

%\bt{\blue{General statement}}. 
The definitions of cryogenic effects are ambiguous. In general, changes observed at cryogenic temperatures, such as variations in electron mobility and carrier freeze-out, can be considered as cryogenic effects. However, these changes do not significantly impact the functionality of FETs. Herein, we solely focus on discussing nonideal effects that may compromise the fundamental operation of FETs. Specifically, this includes addressing the kink effect and hysteresis mentioned in Ref.~\onlinecite{IncandelaJOEDS2018}.

%\bt{\blue{Kink effect}}. 
The kink effect refers to a sudden increase in $I_{\rm DS}$ when a MOS operates in the saturation region with a high $V_{\rm DS}$. This phenomenon was elucidated in Ref.~\onlinecite{Reilly_npjQI_2015}, which attributes it to the generation of electron-hole pairs when $I_{\rm DS}$ reaches a certain threshold. However, the kink effect is exclusively observed in well-established CMOS technology nodes. According to Ref.~\onlinecite{IncandelaJOEDS2018}, this effect is only observed in n-type MOSFETs fabricated using mature 0.16-$\mu m$ CMOS technology with thick oxide layers. The underlying cause can be explained by MOS scaling, where both the supply and gate insulator thickness scale proportionally to maintain a constant vertical electrical field. Moreover, nanometer-scale processing involves higher substrate doping levels. These alterations result in supply voltages below 1.2V, effectively preventing the occurrence of the kink effect as the acquired energy remains lower than the silicon bandgap~\cite{IncandelaJOEDS2018}. Based on these findings from Ref.~\onlinecite{IncandelaJOEDS2018}, it can be inferred that FETs based on 40-nm processing or more advanced technologies will not exhibit the kink effect.

%\bt{\blue{Hysteresis}}. 
The next phenomenon is hysteresis, which refers to the observed discrepancies in the $I_{\rm D}-V_{\rm GS}$ diagram when sweeping the gate voltage $V_{\rm GS}$ forward and backward (with a fixed voltage bias $V_{\rm DS}$). Hysteresis is a significant concern in well-established processing technology, as indicated by various studies~\cite{CretenISSCC2007,Dierickx1988,Ronald2003}. However, it should be noted that only n-type MOSFETs, mentioned in the previous paragraph, exhibit hysteresis. Specifically, the backward threshold voltage is merely 0.22 V, which is lower than the forward threshold voltage of 0.4 V (with a bias voltage $V_{\rm DS}$ set higher than 1.7 V). The subsequent explanation provides insights into this phenomenon. Firstly, when the gate voltage $V_{\rm GS}$ surpasses the threshold voltage, avalanches caused by impact ionization generate a substantial current. Secondly, as the current reaches its maximum value, there is a decrease in threshold voltage due to the kick effect. Finally, while sweeping back the gate voltage $V_{\rm GS}$ , previously occurred impact ionization results in a threshold voltage of 0.22 V~\cite{IncandelaJOEDS2018}. This explanation establishes a connection between hysteresis and kink effect and suggests that FETs based on CMOS tech nodes of 40 nm or more advanced technologies are unlikely to exhibit significant hysteresis.

\subsubsection{Modeling and Validation}\label{section_3_2_2}

%\bt{\blue{General statement}}. 
The following subsection is dedicated to the modeling and validation tests of Field-Effect Transistors (FETs), structured as follows. The initial three sections delve into various types of FET models, including physics-based analytical models, compact semiempirical models, and RF models (small signal model). These sections elaborate on the modeling procedures for each respective model type. The concluding section of this subsection focuses on validating the aforementioned models.

\paragraph*{\textbf{Physics-based Analytical Models}}

%\bt{\blue{General statement and pros and cons of those models}}. 
Physics-based analytical models, as the fundamental frameworks, accurately depict the behavior of FETs based on well-established processing technologies. However, in nano processing technologies, these models often become cumbersome or inadequate due to second-order effects and other nonideal phenomena such as velocity saturation, subthreshold effect, lattice defect, and interface scattering\cite{Physics_of_Semi_2006}. Nevertheless, our focus remains on discussing physics-based models to gain a comprehensive understanding of the underlying causes behind various cryogenic phenomena.

%\bt{\blue{Concepts of Semiconductor Physics and validation of Maxwell-Boltzmann distribution}}. 
The physics of a semiconductor device relies on the electronic band structure and thermodynamic distributions, such as the Fermi-Dirac distribution and its approximation, the Maxwell-Boltzmann distribution. These distributions enable us to calculate the mobile carrier density as it varies with temperature. In the case of a p-type semiconductor based on phosphorus-doped silicon, the charge density consists of three components: electron density ($n$), hole density ($p$), and acceptor charge density ($N_A^-$). By utilizing the Fermi-Dirac function expressed as $f(E)=1/[1+\exp(E-E_F)/(kT)]$, where $E_F$ represents the Fermi energy, we can derive expressions for $n$ and $p$ as follows
\begin{equation}
\begin{aligned}
&n = \int_{E_{\rm c}}^{E_{\rm c}^\prime} f(E)N(E)dE \\ &p = \int_{E_{\rm v}^\prime}^{E_{\rm v}}(1-f(E))N(E)dE
\end{aligned},
\end{equation}
where the $E_{\rm c}$ and $E_{\rm c}^{\prime}$ means the lower limit and the upper limits of the conduction band respectively, $N(E)$ is the state density of electron, $E_{\rm v}^{\prime}$ and $E_{\rm v}$ are the upper and lower limits of valance band. The expression of $N_{\rm A}^{-}$ is as follows
\begin{equation}
N_{\rm A}^{-}= N_{\rm A} \frac{1}{1+g_{\rm A}\exp(\frac{E_{\rm A}-E_{\rm F}}{kT})},
\end{equation}
the multiplier of the doping concentration $N_{\rm A}$ also has the form of the Fermi-Dirac function but with an additional degenerate factor $g_{\rm A}$, and the factor $g_{\rm A}=4$ in phosphorus-doped silicon due to the band structure and electron spin~\cite{HughesNature1967}. At RT, if $E-E_{\rm F}$ is much larger than $kT$, then the Fermi-Dirac function could be replaced by Boltzmann function $exp(-(E-E_{\rm F})/kT)$ approximately, and the semiconductor satisfying this approximation is the so-called non-degenerate semiconductor and electron density can be expressed analytically as $n=N_{\rm c} \exp(-\frac{E_{\rm c}-E_{\rm F}}{k T})$, where $N_{\rm c}$ is the effective state density of conduction band. However, whether a semiconductor is degenerate or not depends not only on $N_{\rm A}$ but also on other physical quantities. Fortunately, the Boltzmann function is still applicable to p-type silicon with $N_{\rm A}=10^{12}$ even at very low temperatures~\cite{BeckersJOEDS2018, BeckersTED2018}. Based on this assumption, the authors developed a physics-based model for long channel devices, which effectively explains the degeneration in SS. Although there are numerous studies addressing the physics-based model~\cite{BohuslavskyiEDL2019, BeckersEDL2020, GhibaudoSSE2020, BeckersJOEDS2018,BeckersTED2018,Beckers2022}, we focus on the methods presented in Ref.~\onlinecite{BeckersJOEDS2018,BeckersTED2018,Beckers2022}. 

%\bt{\blue{Modeling steps of Ref.32}}. 
A comprehensive mathematical derivation is presented in Ref.~\onlinecite{BeckersTED2018}. The study initiates by considering the 1-D Poisson equation based on band structure to determine the mobile charge density per unit ($Q_{\rm m}$) and establish the $|Q_{\rm m}|$-$V_{\rm GB}$ relationship. Subsequently, the current $I_{\rm DS}$ and its corresponding $I_{\rm DS}$-$V_{\rm GB}$ relation are derived using a drift-diffusion transport model. A detailed elucidation follows.

%\bt{\blue{Poisson equation}}. 
The Poisson equation could be derived from the Maxwell equation $\nabla \cdot \mathbf{E} = \rho/\varepsilon$ and $\mathbf{E} = -\nabla\psi$. In the channel and substrate of a MOSFET, the one-dimension approximation is used. According to the charge density in p-type silicon, the Poisson equation is written as 
\begin{equation}
\frac{\partial^2 \psi(y)}{\partial y^2} = -\frac{q}{\varepsilon}(-n+p-N_{\rm A}^-),
\end{equation}
where the $\psi \triangleq (E_{\rm F}-E_{\rm i})/q$ is a function of the depth ($y$) as shown in Fig.~\ref{fig3}(c), and $\varepsilon$ is the silicon permittivity. Through mathematical transformation, the Poisson equation could be rewritten as
\begin{equation}
\frac{\partial^2 \psi(y)}{\partial y^2}=-\frac{q}{\varepsilon}\left( -n_ie^{\frac{\psi-V_{\rm ch}}{U_{\rm T}}} + n_ie^{-\frac{\psi}{U_{\rm T}}}-N_{\rm A}^{-}\right),
\end{equation}
where the $V_{\rm ch}$ is the transversal channel voltage due to the non-zero voltage $V_{\rm DS}$, and $U_T\triangleq \frac{kT}{q}$ is the thermal voltage. Multiply $\partial\psi/\partial y$ at both sides then take the integral from the surface of the device to the end of the substrate. The vertical electric field at the surface $\mathbf{E}_s$ is obtained. Combined with the Gauss theorem $Q_{sc}=-\varepsilon \mathbf{E}_s$ (where $Q_{sc}$ is the total charge density in the substrate) and subtract the fixed charge density (where the charge-sheet and fully depletion approximations~\cite{Yannis1987} is used) the mobile carrier density $Q_m$ is written as
\begin{equation}\begin{aligned}
&Q_{\rm m} =-\varepsilon\sqrt{\frac{2qn_{\rm i} U_{\rm T}}{\varepsilon}\left(e^{\frac{\psi_{\rm s}-V_{\rm ch}}{U_{\rm T}}}-e^{\frac{\psi_{\rm b}-V_{\rm ch}}{U_{\rm T}}}\right)} \\ &\overline{-\frac{2qN_{\rm A}}{\varepsilon}\left[\psi_{\rm s}-\psi_{\rm b}-U_{\rm T}\ln \frac{f_{\rm s}(E_{\rm A})}{f_{\rm b}(E_{\rm A})}\right]} \\ &+\varepsilon \sqrt{\frac{2qN_{\rm A}}{\varepsilon}\left[\psi_{\rm s}-\psi_{\rm b}-U_{\rm T}\ln \frac{f_{\rm s}(E_{\rm A})}{f_{\rm b}(E_{\rm A})}\right]}
\end{aligned},\end{equation}
where $f_{\rm s}(E_A)$ and $f_{\rm b}(E_A)$ are still the Fermi-Dirac function with potential $\psi$ equal to $\psi_s$ and $\psi_b$ correspondingly ($\psi_s$ and $\psi_b$ are the potential at the surface and the end of the substrate). Then the charge density $Q_m$ is connected with gate voltage $V_{\rm GB}$ through expression $V_{\rm GB}=V_{\rm FB}+\varepsilon\mathbf{E}_{\rm s}/C_{\rm ox}+(\psi_{\rm s}-\psi_{\rm b})$ and potential $\psi_{\rm s}$ (potential $\psi_{\rm b}$ is a constant obtained by solving the neutrality equation $p=N_A^{-}$ in substrate).

%\bt{\blue{Drift-diffusion model}}. 
The authors considered the drift-diffusion model in saturation region with form
\begin{equation}
I_{\rm DS}=-\frac{W}{L}\int_{\psi_{\rm s,S}}^{\psi_{\rm s,D}}\mu_{\rm n}Q_{\rm m} d\psi +\frac{W}{L}\int_{Q_{\rm m,S}}^{Q_{\rm m,D}}\mu_{\rm n} U_{\rm T} dQ_{\rm m},
\end{equation}
where $\mu_n$ is the Mobility of electron, the index $S$/$D$ means source/drain, for $Q_m$, we can simply take $V_{\rm ch}$ to be $0$ or $V_{\rm DS}$. In this equation, the hole density is neglected. Assuming the linearization of charge density~\cite{SalleseSSE2003}, the current $I_{\rm DS}$ is expressed as
\begin{equation}
I_{\rm DS}=\frac{W}{L} \mu_{\rm n} \left[ -\frac{Q_{\rm m,D}^2-Q_{\rm m,S}^2}{2mC_{\rm ox}} +U_T(Q_{\rm m_D}-Q_{\rm m,S}) \right],
\end{equation}
where the $C_{\rm ox}$ is the capacitance of the oxide layer per unit area, the constant factor $m\triangleq \partial(Q_m/C_{\rm ox})/\partial \psi_s$. Substitute the expression of $Q_m$ into the equation then $I_{\rm DS}$ only depends on potential $\psi_s$. Besides, the $V_{\rm GB}$ is also a function of $\psi_s$ due to the mentioned equation $V_{\rm GB}=V_{\rm FB}+\varepsilon\mathbf{E}_{\rm s}/C_{\rm ox}+(\psi_{\rm s}-\psi_{\rm b})$. In total, a set of parameter equations for $I_{\rm DS}$ and $V_{\rm GB}$ is established. By changing the parameter $\psi_{\rm s}$, $I_{\rm DS}$-$V_{\rm GB}$ diagram could be drawn. 

%\bt{\blue{Interface traps}}. 
Next, we present the interface traps discussed in~\cite{BeckersJOEDS2018,BeckersTED2018}, which play a crucial role in characterizing the degradation of SS. These interface traps are incorporated into flat-band voltages $V_{\rm FB}$ through the equation $V_{\rm FB}=\phi_{\rm ms}-Q_{\rm it}/C_{\rm ox}$, where $\phi_{\rm ms}$ represents the work function difference between metal and semiconductor, and $Q_{\rm it}$ denotes the trap charge density. The presence of interface traps can be interpreted as discrete or continuous acceptor levels within the band gap equivalently. Therefore, $Q_{\rm it}$ can be determined either by summing over all discrete energy states using $Q_{\rm it}=-q \sum_j N(E_{{it,j}})f_s(E_{{it,j}})$ or by integrating over a continuous energy distribution using $Q_{{it}}=-q \int N(E_{{it}})f_s(E_{{it}}) dE_{{it}}$.

%\bt{\blue{Approxiamtion in Ref.41}}. 
In Ref.~\onlinecite{BeckersJOEDS2018}, a body-partitioning approach was developed to simplify the lengthy equation of charge density $Q_m$. This partitioning scheme is based on the relative relationships among $E_F$, $E_A$, and $E_c-3kT$, specifically considering the bulk region (where the band is flat), freeze-out layer ($E_{F}\leq E_{A}$, where the band is bent), ionized layer ($E_A\leq E_F<E_c-3kT$), and inversion layer ($E_F>E_c-3kT$). Unlike the rigorous treatment in $p-n-N_A^-$, as described in~\cite{BeckersTED2018}, certain types of ionization are neglected in some regions. Specifically, no ionization is assumed in the bulk region. In the freeze-out layer, incomplete ionization of acceptor $N_A^-$ is assumed while ignoring electron concentration ($n$). In the ionized layer, complete ionization is adopted while neglecting hole concentration ($p$). Finally, only electron concentration ($n$) is considered in the inversion layer. The derivation of charge density follows a similar approach as described in Ref.~\onlinecite{BeckersTED2018}; therefore, we will not elaborate on it here. Reference~\onlinecite{Beckers2022} primarily discusses SS modeling using band tail effects instead of interface traps; however, we will not delve into its details here. Nevertheless, this method holds potential for modeling without introducing ambiguity associated with interface trap levels.

\paragraph*{\textbf{Compact Models}}

%\bt{\blue{General statement}}. 
Compact models for nanometer devices are often characterized by a set of parameters that lack physical insights, despite being empirically derived from characterization results. However, these models exhibit high accuracy due to their ability to depict second-order effects. In this context, compact models can be categorized into three types for simulating DC properties~\cite{IncandelaJOEDS2018,BeckersJOEDS2018,SinghEDL2022}, mismatch and self-heating effects~\cite{HartESSDERC2018,HartJEDS2020,HartJEDS2021}, and those based on artificial neural networks~\cite{HartWOLTE2021}.

%\bt{\blue{Modeling process of DC parameters}}. 
The DC properties of FETs, as mentioned in the characterization section, primarily encompass the $I_{\rm DS}$-$V_{\rm GS}$ relationship and the $I_{\rm DS}$-$V_{\rm DS}$ relationship with fixed biased voltages $V_{\rm DS}$ and $V_{\rm GS}$ respectively. When employing compact models, the modeling process revolves around parameter fitting based on characterization results. In Ref.~\onlinecite{IncandelaJOEDS2018}, the MOS11~\cite{Ronald2003} and PSP~\cite{GennadyTED2006} models were selected, while a simplified EKV model~\cite{Christian2006, ChristianSSCM2017, ChristianSSCM2017_2} was adopted in Ref.~\onlinecite{BeckersJOEDS2018}. Additionally, commonly used models such as the BSIM models~\cite{Chua2013} are available, and Reference ~\onlinecite {SinghEDL2022} presents a modified BSIM-CMG model for FinFETs. We summarize the modeling procedure introduced in those references as follows. In total, based on the relationship between ambient temperature $T$ and the critical temperature $T_c$ (the temperature at which the factor $n$ in SS undergoes significant changes), the parameters of the model are fitted separately. The approach employed in both parts may be similar~\cite{IncandelaJOEDS2018} or different~\cite{SinghEDL2022}, but the parameters related to SS are all fitted first, followed by fitting important temperature-dependent parameters such as $V_{\rm TH}$ and $g_{\rm Dm}$. This procedure aims to minimize deviations between RT and cryogenic temperatures, with an adjustment made to obtain higher accuracy when $T<T_c$, as described in Ref.~\onlinecite{SinghEDL2022}.

%\bt{\blue{Modeling process of mismatch}}. 
We discuss the models of mismatch mentioned in the previous subsection based on Ref.~\onlinecite{HartESSDERC2018,HartJEDS2020}, where the Croon and Pelgrom models~\cite{Marcel1991,Jeroen2005} are adopted. The essence of both models lies in extracting the area-scaling parameters $A_{TH}$, $A_{\beta}$, and $A_{SS}$ at cryogenic temperatures. These parameters can characterize the standard deviations of $V_{\rm TH}$, $\beta$ (current factor), and SS according to Pelgrom's scaling law
\begin{equation}
\sigma_{\Delta V_{\rm TH}} =\frac{A_{\rm TH}}{\sqrt{WL}}\ \sigma_{\Delta \beta/\beta} =\frac{A_{\beta}}{\sqrt{WL}}\ \sigma_{\rm \Delta SS/SS} =\frac{A_{\rm SS}}{\sqrt{WL}}.
\end{equation}
With those parameters, the standard derivation of drain current $I_{\rm D}$ could be obtained
\begin{equation}
\sigma^2_{\Delta\log I_{\rm D}}=\frac{1}{\ln(10)^2}\left[\sigma^2_{\Delta\beta/\beta} + (\frac{\bar{g}_{\rm m}}{\bar{I}_{\rm D}})^2 \sigma^2_{\Delta V_{\rm TH}}\right],
\end{equation}
where the $\bar{g}_m$, $\bar{I}_D$ are the mean value of transconductance and drain current. The extraction procedure for the standard derivations and mean values can be effectively implemented through experimental methods, which do not require sophisticated techniques.

%\bt{\blue{Modeling process of self-heating effect}}. 
The self-heating effect is also important, but not discussed widely in cryogenic FETs. Here, we introduce the modeling of self-heating in Ref.~\onlinecite{HartJEDS2021} as an example. The model of this work is similar to~\cite{TriantopoulosTED2019}, which is described by the thermal resistance $R^*_{\rm th} \triangleq d\Delta T_{\rm chan}/d P_{\rm H}$ (where the $T_{\rm chan}$ is the increased temperature compared with the ambient temperature, and $P_H$ is the power supplied by the heater on chip). As a function of $T_{\rm chan}(\Delta T)$, $R^*_{\rm th}$ could be fitted by empirical analytical formula. With $R^*_{\rm th}$, increased temperature in channel $\Delta T$ which is a function of provided power $P$ and ambient temperature $T_{\rm amb}$ can be expressed by the equation below implicitly.
\begin{equation}
P=\int_0^{\Delta T}\frac{d\Delta T^{\prime}}{R^*_{\rm th} (T_{\rm amb}+\Delta T^{\prime})}.
\end{equation}

%\bt{\blue{Artificial Neural Networks Model}}. 
The utilization of Artificial Neural Networks (ANNs) for modeling Field-Effect Transistors (FETs) is discussed at last. As widely adopted tools across various research domains, ANN models eliminate the need for equation development or user-defined parameter extraction processes~\cite{HartWOLTE2021}. In Ref.~\onlinecite{HartWOLTE2021}, ANNs are trained with a diverse range of input data including drain current variations with different geometries, ambient temperature, and bias conditions to automatically obtain DC models. This same methodology can be extended to RF models by substituting the samples with S-parameter characterization results.

\paragraph*{\textbf{RF Models}}

%\bt{\blue{General statement}}. 
In most cases, the RF models of FETs primarily consist of small signal models. For instance, Fig.~\ref{fig3}(d) illustrates a small signal model of a 22-nm FDSOI MOSFET. The essence of the modeling procedure lies in extracting capacitance and resistance for small signal models, while transconductances are obtained from DC characterization results. Furthermore, we discuss the variations between RT and cryogenic temperatures during this process. In this review, we consider Refs.~\onlinecite{ChakrabortyIEEE2021,TangJEDS2022,HanESSDERC2022} as representative examples of RF models.

%\bt{\blue{Modeling Procedure}}. 
We first consider Ref.~\onlinecite{HanESSDERC2022}, where the small signal models are depicted in Fig.~\ref{fig3}(d). In this study, for simplification purposes, the junction capacitance and substrate resistance are replaced by the back-gate resistance $R_B$ and box capacitances. Both the real and imaginary parts of Y parameters (with FET biased at saturation region) can be expressed as polynomial forms~\cite{Chalkiadaki2015} such as $a_0 + a_1\omega + a_2\omega^2$. Subsequently, an iteratively re-weighted least-squares (IRLS) method is employed to extract capacitances and resistances in these models. Other works like Refs.~\onlinecite{ChakrabortyIEEE2021,TangJEDS2022} utilize Smith charts for parameter extraction. Generally, the procedure for extracting RF parameters can be divided into three steps: 1) Deriving the relationship between Z/Y parameters and RF parameters using phasor analysis; 2) Converting S parameters to Z/Y parameters; 3) Fitting the RF parameters.

%\bt{\blue{Results contracted with RT}}. 
The results from all three references demonstrate that gate capacitances exhibit negligible sensitivity to temperature variations, whereas junction capacitances exhibit a reduction as the temperature decreases. Specifically, junction capacitances decrease by approximately 8\% from room temperature (RT) to 6K. Additionally, the resistances of the source and drain experience a decline due to degenerated phonon scattering.

\paragraph*{\textbf{Validation of Models}}

%\bt{\blue{Validation Analysis}}. 
The validation and comparison of three models are included in this section. The physics-based model proposed in Ref.~\onlinecite{BeckersJOEDS2018, BeckersTED2018} presents an analytical formula to accurately describe the DC characteristics of FETs, particularly for long channel devices such as $W/L=3\mu m/1\mu m$ 28-nm bulk MOSFET. However, its applicability is limited due to neglecting effects in nanometer devices. Additionally, the computational complexity of the analytical formula hinders its application in large-scale circuit simulations. In contrast, the compact models introduced in Ref.~\onlinecite{IncandelaJOEDS2018,SinghEDL2022} leverage mature nanometer commercial models and are more suitable for simulating large circuits. Although a discrepancy of about 10\% between model predictions and experimental data is observed in~\cite{IncandelaJOEDS2018}, the model predictions presented in~\cite{SinghEDL2022} align well with experimental data. Furthermore, the ANN-based model proposed by Hart et al.\cite{HartWOLTE2021} also demonstrates accurate prediction of DC properties; however, it exhibits degradation in performance within the middle inversion region due to limitations associated with continuity properties inherent to ANN functions. Overall, a key limitation of compact models lies in their dependence on FET geometry size parameters; specifically, extracting finer parameters necessitates numerous samples with varying sizes (e.g., $W/L$, finger number). Finally, the RF model developed by Chakraborty et al.\cite{ChakrabortyIEEE2021,TangJEDS2022,HanESSDERC2022} effectively captures experimental RF responses. Similar to compact models describing DC properties, small signal RF models also exhibit geometry-dependence; fortunately though, most RF parameters demonstrate low sensitivity to ambient temperature.

\subsection{Designs and Performances of Cryo-CMOS ICs}\label{section_3_3}

%\bt{\blue{Introduction}}. 
After the publication of the proposals in 2017 (Ref.~\onlinecite{Charbon_ISSCC_2017}), extensive research has been conducted to develop prototypes of Cryo-CMOS integrated circuits (ICs). In this section, we will provide a comprehensive review of recently developed Cryo-CMOS prototype chips from both design and performance perspectives.

\subsubsection{Pulse Modulator/Control Module}\label{section_3_3_1}

%\bt{\blue{General statement}}
As outlined in As described in Section~\ref{section_2_2} and Section~\ref{section_3_1}, the pulse modulator serves a similar purpose to that of a transmitter, with its primary function being the generation of envelope-modulated single-frequency microwave signals for qubit manipulation on the Bloch sphere. This module, as an integral component of quantum-classical interfaces, has been extensively investigated. The review of this module can be categorized into three sections based on different modulation architectures: 1. Basic IQ modulation, 2. I/Q single sideband modulation (SSB), and 3. Polar modulation.

\paragraph*{\textbf{Direct Up-conversion Modulation}}

%\bt{\blue{Statement of the References /prototypes discussed in this subsection}}. 
Due to the unique nature of qubit control signals discussed in Section~\ref{section_2_2}, the necessity of arbitrary wave generation (AWG) function in ASICs may be obviated. Moreover, circuit designs need to be simplified for minimizing power consumption in a cryostat. In this section, we focus on Google's research (~\cite{YooISSCC2023, BardinISSCC2019}) that achieves pulse modulation through direct up-conversion.

\begin{figure*}
\centering
\includegraphics[width=1\linewidth]{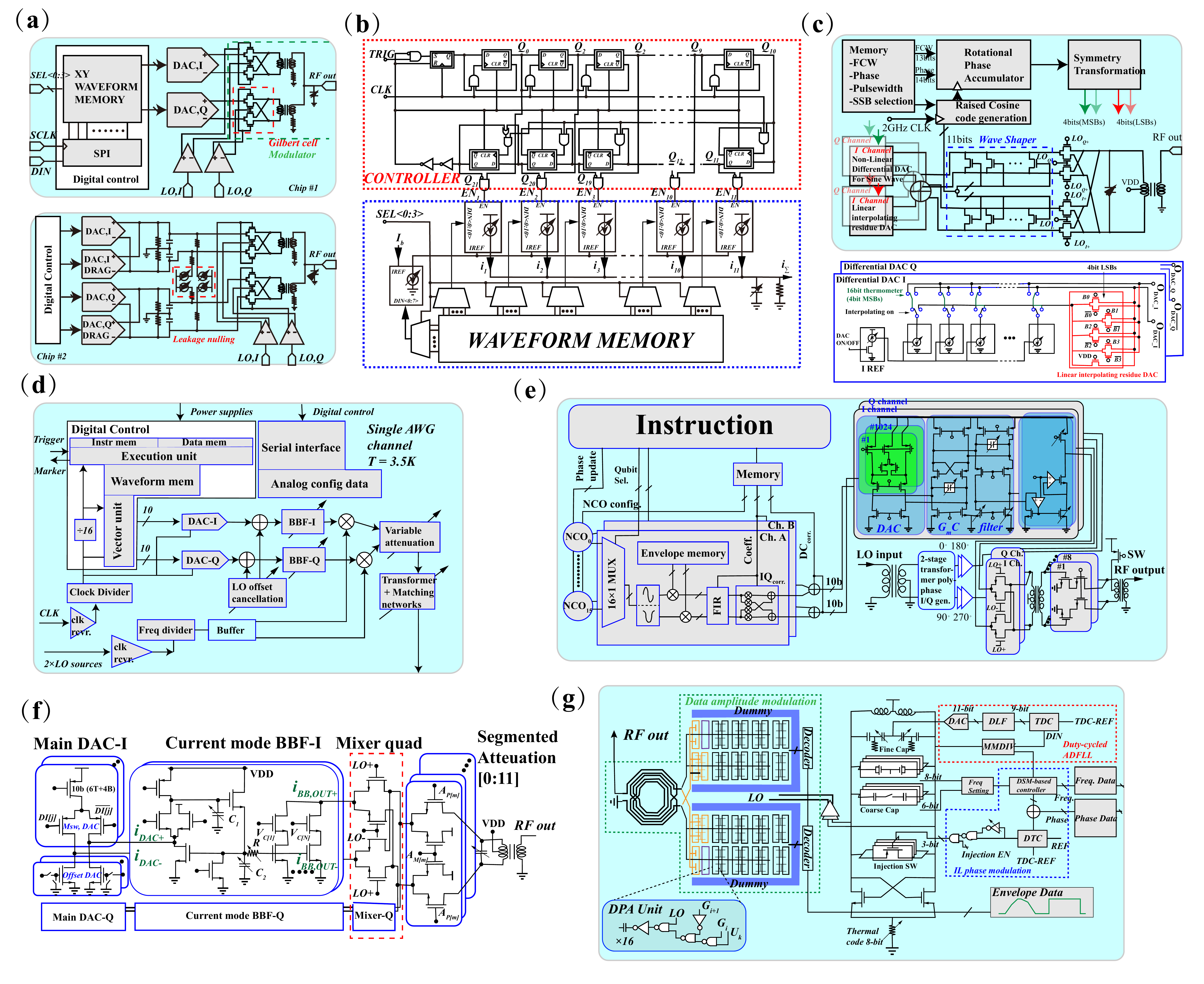}
\caption{\bt{System structures of modulator prototypes.} (a) System structure of Google's works. (chip\#1,2 corresponds to~\cite{BardinISSCC2019, YooISSCC2023} separately.) (b) Interconnection relation of the digital part and DACs in~\cite{BardinISSCC2019}. (c) System structure and schematic of DAC in~\cite{KangISSCC2023, KangISSCC2022, KangVLSI2021}. (d) System architecture of IBM's work~\cite{FrankISSCC2022}. (e) Chip in Reference~\cite{ParkJSSC2021}. (f) Gate-level structure in~\cite{FrankISSCC2022}. (g) Chip in Reference~\cite{GuoISSCC2023}. Panel (a) is from Refs.~\cite{BardinISSCC2019, YooISSCC2023}, (b) is from Ref.~\cite{BardinISSCC2019}, (c) is from Refs.~\cite{KangISSCC2023, KangISSCC2022, KangVLSI2021}, (e) is from Ref.~\cite{ParkJSSC2021}, (f) is from Ref.~\cite{FrankISSCC2022}, (g) is from Ref.~\cite{GuoISSCC2023}. }\label{fig4}
\end{figure*}

%\bt{\blue{Brief Introduction of Google's chips}}. 
The ICs utilized in Google's research are based on the superconducting transmon platform. A notable distinction from conventional transmitter architecture is that the IF signals are  proportional to the function $s(t)$ described by Equation (2). The modulators and digital control structures presented in Fig.~\ref{fig4}(a) of ~\cite{BardinISSCC2019, YooISSCC2023} illustrate their design implementation. In~\cite{BardinISSCC2019}, each block follows a similar design approach as outlined in~\cite{YooISSCC2023}; however, two additional DACs have been incorporated to facilitate the realization of XY control through the DRAG scheme.

%\bt{\blue{Detailed discussion of Google's chips}}. 
In Fig.~\ref{fig4}(a), the Digital component in~\cite{BardinISSCC2019} comprises waveform memory and a DAC controller based on a shift register, as depicted in Fig.~\ref{fig4}(b). The interconnections between these components are illustrated. The memory blocks have the capacity to store 16 different types of envelopes, and a 4-bit binary signal $SEL\langle 0:3\rangle$ is utilized to select the desired envelope waveforms through MUXs. The DAC block consists of 11 8b current mode DACs, each employing an 8-bit binary-weighted current mirror circuit with a common reference current $I_{\rm N}$. The controller's registers (totaling 11) are connected to the DAC, enabling sequential activation of sub-DACs numbered from 1 to 11 during the first set of clock periods upon detection of the trigger signal; subsequently, they are disabled inversely over the next set of clock periods. Consequently, a wide envelop signal spanning across 22 periods is generated and then converted into differential form before being filtered. For mixer operation, a quadrature double-balanced Gilbert-cell active mixer is employed along with on-chip transformers for converting input I/Q LO currents into differential forms. Additionally, in~\cite{YooISSCC2023}, two extra DACs are included along with gate-level instruction sets and sequencers within the digital part to facilitate algorithm execution up to 512 steps; furthermore, an LO leakage nulling block follows after filtering stage. We now proceed to explain how this modulator implements the DRAG scheme by referring back to Equation (4) in Section~\ref{section_2_2}, which can be expanded as
\begin{equation}
\begin{aligned}
f(t)&=(S_{\rm M}(t)\cos\phi-S_{\rm D}(t)\sin\phi)\cos(\omega_{\rm 01}t) \\ &+ (-S_{\rm M}(t)\sin\phi+S_{\rm D}(t)\cos\phi)\sin(\omega_{\rm 01} t)
\end{aligned},
\end{equation}
where the $S_M(t)$ is equivalent to the $s(t)$ in Equation (4), and the $S_D(t)$ corresponds to the $\lambda \dot{s}(t)/\alpha$. Since all the referenced works~\cite{YooISSCC2023, BardinISSCC2019} employ the direct up-conversion architecture ($\omega_{\rm LO} = \omega_{01}$), it is possible to obtain the DRAG signals by generating currents $S_M(t)\cos\phi$, $–S_D(t)\sin\phi$, $-S_M(t)\sin\phi$, and $S_D(t)\cos\phi$ in channels I, I$_{\rm DRAG}$, Q, and Q$_{\rm DRAG}$ respectively.

\paragraph*{\textbf{SSB Modulation}}

%\bt{\blue{General statemsnt}}. 
The SSB modulation architecture is extensively employed in this domain. This section's review primarily focuses on several notable works, including IBM's study~\cite{FrankISSCC2022}, Stefano Pellerano et al.'s research~\cite{PatraISSCC2020,XueNature2021,ParkJSSC2021}, and Kiseo Kang et al.'s investigations~\cite{KangISSCC2023, KangISSCC2022, KangVLSI2021}. These studies incorporate more intricate digital control components compared to the previously discussed works and also utilize non-zero IF frequencies. However, there exist distinctions among these works; specifically, IBM's work~\cite{FrankISSCC2022} adopts the AWG function while digital direct synthesis (DDS) is employed in ~\cite{PatraISSCC2020,XueNature2021,ParkJSSC2021,KangISSCC2023, KangISSCC2022, KangVLSI2021}.

%\bt{\blue{Chips of Delft/Intel}}. 
We started with~\cite{PatraISSCC2020,XueNature2021,ParkJSSC2021}, These studies share a consistent framework in terms of the modulator block and digital control block design. In particular, we focus on the work in Ref.~\onlinecite{ParkJSSC2021} and present its system structure in Fig.~\ref{fig4}(e). The left side of the figure illustrates the digital part, where an instruction controller provides phase codes to 16 numerical control oscillators (NCOs), while a qubit selection code is provided to a $16\times 1$ multiplexer (MUX). The memory block is programmable and stores frequency configurations of NCOs, I/Q correction coefficients, and DC offset codes. Similarly, the envelope memory serves similar functions as well. During operation, the MUX selects signals generated by NCOs represented as 10-bit codes which are then converted into 16-bit sine and cosine codes through look-up tables (LUTs). After passing through finite impulse response (FIR) filters, these sine and cosine codes are multiplied by coefficients stored in memory before being added together. Finally, both channel A and channel B's resulting codes are combined before entering DACs to provide control signals for two qubits. On the right side of Fig.~\ref{fig4}(e), we illustrate the transistor-level system structure of this modulator design where 10-bit 2.5GS/s current steering DACs convert outputted 10-bit codes into analog domain signals that undergo smoothing via second-order Gm-C reconstruction filters followed by amplification using variable gain amplifiers (VGAs). While still based on Gilbert cells, subsequent circuits after mixers function as filters.

%\bt{\blue{IBM's chips}}. 
The work conducted by IBM~\cite{FrankISSCC2022} is illustrated in Fig.~\ref{fig4}(d). The digital block represents a domain-specific processor capable of executing 32-bit fixed-point instructions. As depicted in Fig.~\ref{fig4}(d), different subblocks within the digital block are controlled by clocks operating at varying frequencies. The fast clock domain, with a nominal frequency of 1-1.125 GHz denoted as $f_{\rm s}$, is allocated to the vector unit for waveform play operations. To minimize power consumption, the slow clock domain operates at 1/16$f_{\rm s}$ and is utilized for program control purposes. This microprocessor adopts a single-cycle microarchitecture comprising instruction fetch, decode, branch resolution, and scalar arithmetic execution stages; its core function involves calculating waveform coefficients. Referring to the DRAG signals $f(t)=\gamma E_I(t)\sin(\omega_{01}t+\phi)+\gamma E_Q(t)\cos(\omega_{01}t+\phi)$ (where $\gamma$ denotes a constant; compared with Equation (5), I and Q components are inverted in~\cite{KangISSCC2023}), substituting the qubit frequency $\omega_{01}$ with $\omega_{IF} + \omega_{LO}$ and expanding the equation yields desired signals that should take on the following forms
\begin{equation}\begin{aligned}
&f(t)=\\&\gamma (E_{\rm I}(t)\cos(\omega_{\rm IF} t +\phi)-E_{\rm Q}(t) \sin(\omega_{\rm IF}t+\phi))\sin\omega_{\rm LO}t +\\& \gamma (E_{\rm I}(t)\sin(\omega_{\rm IF}t+\phi) + E_{\rm Q}(t) \cos(\omega_{\rm IF}t+\phi)\cos\omega_{\rm LO}t
\end{aligned},
\end{equation}
the coefficients of $\sin(\omega_{\rm LO}t)$ and $\cos(\omega_{\rm LO}t)$ represent the desired output signals in the I and Q channels, denoted as $V_I(t)$ and $V_Q(t)$, respectively. In~\cite{ChakrabortyJSSC2022}, $V_I(t)$ and $V_Q(t)$ are defined as $V_I(t)=\alpha U_I(t) -\beta U_Q(t)$, $V_Q(t)=\alpha U_Q(t) + \beta U_I(t)$, where $U_I(t)=E_I(t)\cos(\omega_{\rm IF} t)-E_Q(t)\sin(\omega_{\rm IF} t)$, and $U_Q=E_I(t)\sin(\omega_{\rm IF} t) + E_Q'(t)\cos(\omega_{\rm IF} t)$. Here, $\alpha = \gamma \cos(\phi)$ and $\beta = \gamma \sin(\phi)$. The determination of these main coefficients is crucial. Returning to the system structure, Fig.~\ref{fig4}(f) presents the transistor-level architecture. To balance differential nonlinearity errors with power consumption considerations, 6-bit thermometer MSB segmentation combined with 4-bit binary LSB segmentation is employed in DAC design. Current mode BBFs~\cite{GianniniISSCC2019} are utilized as continuous-time filters and amplifiers. The mixer configuration remains consistent with previous works reviewed herein. Prior to reaching the final transformer stage, 12 segments of attenuators are incorporated.

\begin{figure*}
\centering
\includegraphics[width=1\linewidth]{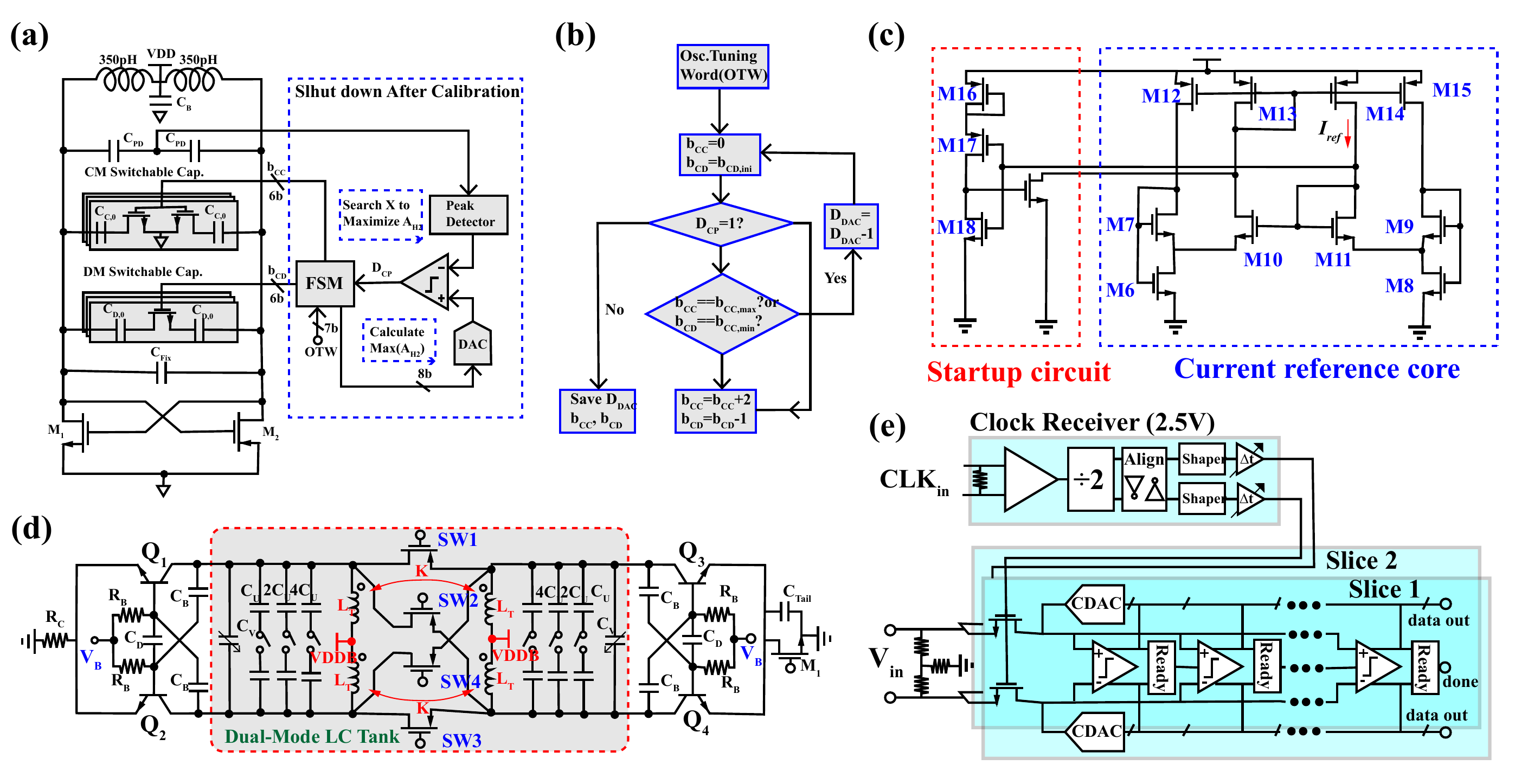}
\caption{\bt{Oscillator, VCO, and ADC.} (a) System structure of oscillator in~\cite{GongISSCC2020}. (b) Calibration procedure of oscillator in~\cite{GongISSCC2020}. (c) Reference current generation circuit in~\cite{HuISCAS2022}. (d) Gate-level structure of VCO in~\cite{PengISSCC2022}. (e) System structure of ADC in~\cite{KieneISSCC2021}. Panels (a) and (b) are from Ref.~\cite{GongISSCC2020}, (c) is from Ref.~\cite{HuISCAS2022}, (d) is from Ref.~\cite{PengISSCC2022}, (e) is from Ref.~\cite{KieneISSCC2021} }\label{fig7}
\end{figure*}

%\bt{\blue{Pohang University's chips}}. 
In contrast to the conventional transmitter architecture, the modulator blocks in Ref.~\onlinecite{KangVLSI2021, KangISSCC2022, KangISSCC2023} employ pulse shaping techniques for generating the envelope waveform. The IF signals are generated by DACs, which consist of both non-linear and linear interpolating components. The system frameworks of the modulator and digital control block are illustrated in Fig.~\ref{fig4}(c). (Since all these ICs adopt the scheme presented in the earliest work~\cite{KangVLSI2021}, we present its framework as described in~\cite{KangVLSI2021}.) The digital block of the modulator comprises four parts: memory, pulse shaping code generation, rotational phase accumulator, and symmetry transform. It utilizes a 13-bit frequency code and a 14-bit initial phase as inputs for the rotational phase accumulator. These values are then passed to the symmetry transform block to generate 4-bit MSBs and LSBs for DAC operation. Additionally, an 11-bit pulse shaping code is generated to achieve desired envelopes. The entire process is controlled by an external clock with a frequency of approximately 2 GHz, as described in~\cite{KangVLSI2021}. In~\cite{ KangISSCC2022}, on-chip PLL generates this clock signal. The schematic diagram of DAC is shown at the bottom of Fig.~\ref{fig4}(c), where each current mirror refers to an identical 8-bit current source; while MSBs are converted into a 16-bit thermal code to drive non-linear DACs, LSBs are provided for linear interpolation section operation. After generating sinusoidal wave pulses in I/Q channels, they undergo filtering and shaping using FETs controlled by pulse shaping codes.The mixer employed remains Gilbert cell type. Following up-conversion, the signals undergo a secondary filtration process to eliminate undesired frequencies. It is noteworthy that in~\cite{KangISSCC2023}, two supplementary I/Q channels and DACs are incorporated similar to~\cite{YooISSCC2023} for implementing the DRAG scheme. However, these four DACs are employed to generate sinusoidal waves with distinct frequencies and initial phases. Consequently, without pulse shaping, the ultimate output signal can effectively achieve the DRAG signal with sinusoidal envelopes.

\paragraph*{\textbf{Polar modulation}}

%\bt{\blue{2023 Tsinghua's chip}}. 
While most modulators are designed based on I/Q modulation, there is also a study~\cite{GuoISSCC2023} that utilizes polar modulation. The system structure of this approach is illustrated in Fig.~\ref{fig4}(g). This architecture can be divided into five blocks, namely memory, injection-locking (IL) phase modulator, duty-cycled all-digital frequency-locked loop (ADFLL), DCO, and amplitude modulation-based switched-capacitor digital power amplifier (DPA). The first three blocks serve the purpose of phase modulation and LO signal generation. Upon the initiation of global signals, the trigger signal activates the FIFO which provides 20-bit frequency control words (FCW) and 10-bit phase control words (PCW) to the IL phase modulator. Subsequently, a delta-sigma modulator generates 14-bit coarse capacitor control words (CCW) to roughly regulate DCO. Simultaneously, the injection EN is enabled and then duty-cycled ADFLL automatically adjusts the frequency of DCO until achieving optimal 11-bit fine CCM with a setup time $t_s<1\mu s$. Finally, only duty-cycled ADFLL switches to the injection locking path for stabilizing both frequency and phase of DCO while obtaining an LO signal (10-bit code) with desired characteristics that enters into DPA. As the LO signal adopts differential form, each part of DPA is segmented into two sections: one consisting of a 6-bit thermal code and another comprising a 4-bit binary code; moreover, every unit within DPA relies on switched capacitors with capacitance value set at 0.5fF as depicted in Fig.~\ref{fig4}(g). Additionally, an extra sub-array employing negative capacitors with a capacity range spanning four bits is incorporated to suppress off-state spur.

\paragraph*{\textbf{DAC}}

%\bt{\blue{DAC in HuISCAS2022}}. 
Finally, we examine a study (Reference~\onlinecite{HuISCAS2022}) on cryogenic DACs as an adjunct to pulse modulator designs. The 8-bit current steering DAC presented in~\cite{HuISCAS2022} employs a 4-bit thermometer-decoded MSB and a 4-bit binary-weighted LSB segmented structure, which bears resemblance to the DAC described in~\cite{FrankISSCC2022}. Notably, this design incorporates an on-chip self-biased current reference, distinguishing it from most cryogenic DACs that still rely on external current references at room temperature due to the unavailability of effective field-effect transistor models~\cite{ZuritaSSCL2020, Rahman2016, RahmanTCS2017}. Fig.~\ref{fig7}(c) illustrates its schematic configuration. The core circuit is based on a symmetric self-biased current source topology with the inclusion of an inverter-type start-up circuit to ensure stable quiescent operation across the entire temperature range. Within the core circuitry, FETs $M_8$-$M_{11}$ are biased within the weak inversion region to maintain nanoampere-level $I_{ref}$; $M_6$ and $M_7$ operate within the moderate inversion region to mitigate mismatch-induced non-linearity; and appropriately sized transistors ($M_{12}$-$M_{15}$) are selected for accurate implementation of current mirroring functionality. Ultimately, characterization results demonstrate that this DAC exhibits minimal power consumption of 13.8$\rm{\mu W}$, a stable output range of 6 mV, and achieves a sampling rate of 140 MS/s at cryogenic temperatures.

\subsubsection{RF Receiver/Readout Module}\label{section_3_3_2}

%\bt{\blue{General statement}}. 
This section provides a comprehensive review of the readout module designs for Cryo-CMOS ICs, encompassing prototype chip designs as well as significant advancements in LNA and ADC research.

\paragraph*{\textbf{Readout prototype chips}}

In this part, we focous on the designs of RF receivers in Ref.~\onlinecite{RuffinoISSCC2021,PrabowoISSCC2021,ParkJSSC2021,KangISSCC2022}, and introduce them separately. 

\begin{figure*}
\centering
\includegraphics[width=1\linewidth]{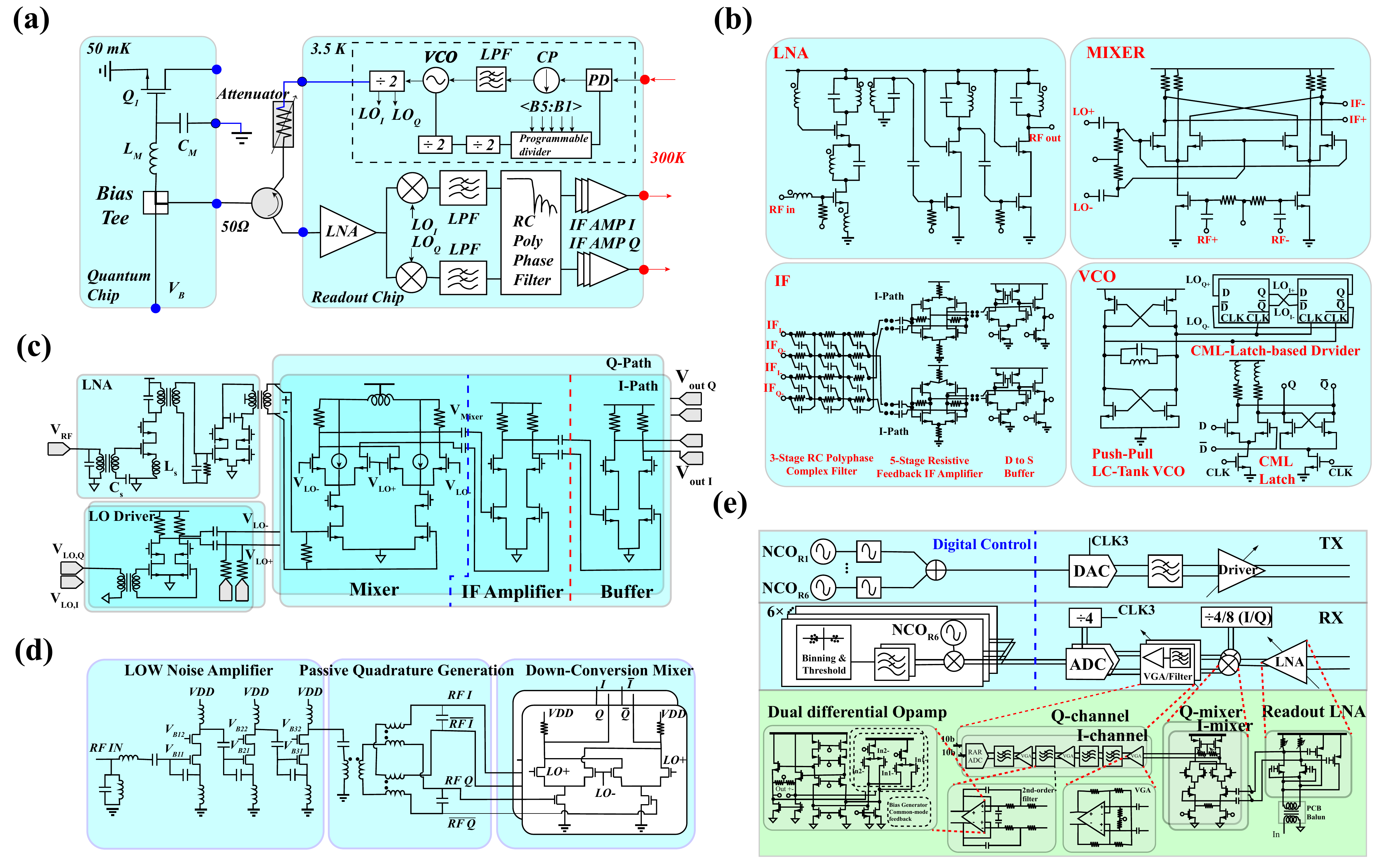}
\caption{\bt{System structure of the RF receiver prototypes.} (a) Transceiver in~\cite{RuffinoISSCC2021}. (b) Block designs in~\cite{RuffinoISSCC2021}. (c) Gate-level system structure in~\cite{PrabowoISSCC2021}. (d) Gate-level system structure in~\cite{KangISSCC2022}. (e) System structure and Block designs in~\cite{ParkJSSC2021}. Panels (a) and (b) are from Ref.~\cite{RuffinoISSCC2021}, (c) is from Ref.~\cite{PrabowoISSCC2021}, (d) is from Ref.~\cite{KangISSCC2022}, (e) is from Ref.~\cite{ParkJSSC2021}.}\label{fig5}
\end{figure*}

%\bt{\blue{Works of Ruffino ISSCC 2021}}. 
The RF receiver system structure depicted in Fig.~\ref{fig5}(a) is presented in~\cite{RuffinoISSCC2021}. In addition to the receiver, this work encompasses an analog charge-pump integer-N PLL and an I/Q receiver module. At the heart of the PLL lies a 12.8 GHz LC VCO, which is regulated by a programmable divider, enabling the VCO output frequency to span from 10.5 GHz to 16.5 GHz. The VCO output signal is divided by a current-mode logic (CML) divider, generating both readout pulses and I/Q components for down-conversion purposes. The transistor-level structures of the CML-Latch-based Driver and VCO are illustrated in Fig.~\ref{fig5}(b). To mitigate significant noise fluctuations at cryogenic temperatures, non-zero IF architecture is employed in RF receivers as shown at the bottom of Fig.~\ref{fig5}(a). The front end comprises a wideband low-noise amplifier (LNA), with its transistor-level configuration displayed in Fig.~\ref{fig5}(b). This design adopts cascaded inductively degenerated common-source stages, where an LC loop replaces conventional drain resistors to optimize noise impedance, achieving a gain of 40 dB within the band ranging from 4.5 GHz to 8.5 GHz.The mixers also employ Gilbert cells similar to those used in modulators.Following the mixer stage,a three-stage RC passive complex polyphase filter(PPF),five-stage IF I/Q amplifier with a gain of 30 dB,and differential-to-single-ended(D/S) buffer are connected successively.The three-stage PPF achieves more than 30 dB image rejection ratio(IRP),while D/S buffer is designed for matching with a 50$\Omega$ output.

%\bt{\blue{Works of Prabowo ISSCC 2021, Kang ISSCC 2022}}. 
Chips in Ref.~\onlinecite{PrabowoISSCC2021, KangISSCC2022} exhibit a similar system-level structure to the one in~\cite{RuffinoISSCC2021}, so we are only concerned with the parts that have significant differences. The transistor-level system structures of~\cite{PrabowoISSCC2021, KangISSCC2022} are shown in Fig.~\ref{fig5}(c) and Fig.~\ref{fig5}(d), respectively. Similar to the previously reviewed works, the mixers still utilize Gilbert cells. The design of LNA in~\cite{PrabowoISSCC2021} is similar to that in~\cite{RuffinoISSCC2021}, except for the addition of a large capacitor $C_s$ after the secondary winding of the transformer; meanwhile, LNA in~\cite{KangISSCC2022} is almost identical to that in~\cite{RuffinoISSCC2021}. The signals after down-conversion are transported directly to the RT electronics without any buffers following them in~\cite{KangISSCC2022}. However, in~\cite{PrabowoISSCC2021}, IF signals pass through two common-source amplifiers after the mixer, where the first one provides gain and serves as a buffer.

%\bt{\blue{Works of Park JSSC 2021}}. 
The RF receiver in Ref.~\onlinecite{ParkJSSC2021} is more comprehensive compared to~\cite{RuffinoISSCC2021}, as it includes the ADC block as well. The system structure is depicted at the top of Fig.~\ref{fig5}(e). The readout pulse generation section utilizes a 2.5GS/s 10-bit DAC, which aims to suppress LO leakage and prevent receiver saturation caused by spurious tones. Prior to the DACs, the digital part comprises of 6 NCOs that enable simultaneous readout of up to 6 qubits through frequency division multiplexing (FDM). In addition to the LNA, mixer, and ADC, the receiver block also incorporates an additional digital component. Within this digital part, both output codes (I and Q) from the DAC along with their five copies are multiplied by their corresponding NCOs and subsequently filtered using digital filters. These processed signals manifest as scatter points on a constellation diagram (I-Q diagram), allowing for determination of $|0\rangle$ or $|1\rangle$ after crossing a predefined threshold value. Returning to the analog portion, transistor-level structures of specific blocks are presented at the bottom of Fig.~\ref{fig5}(e). The LNA employs negative and positive dual-feedback loops~\cite{WooISSCC2009}, enabling achievement of a simulated gain of 27 dB with noise figure below 3 dB (RT). The mixer section remains consistent with previous works but incorporates an additional tunable gain ranging from 2 to 13 dB. Following the mixer stage is a tunable eighth-order active VGA and filter chain; each VGA unit can provide adjustable gain between 5.4 and 18 dB while individual filters allow bandwidth adjustment from 80 to 200 MHz.The final component is a SAR ADC operating at a sampling rate of400 MS/s with effective resolution reaching7 .5 bits; however,the detailed structure is not provided in ~\cite {ParkJSSC2021}.

\paragraph*{\textbf{LNA and ADC}}
Although the designs of blocks like LNA and ADC have been metioned in previous part, there are also attractive works about those blocks. Here we introduce LNA in~\cite{Qu2019} and ADC in~\cite{KieneISSCC2021} as examples.

%\bt{\blue{LNA in Qu2019}}. 
Unlike the RF modulators, the LNA in~\cite{Qu2019} is based on a 70 nm GaAs metamorphic high electron mobility transistor (mHEMT). The structure still incorporates three common-source stages. The first stage utilizes inductive source degeneration to enhance the real part of impedance, while smaller transistors are employed in the second and third stages for gain equalization. Experimental results demonstrate a minimum noise temperature of 8.7 K, achieving a gain above 32 dB within a bandwidth of 6.0-7.0 GHz, with a power consumption of 30 mW.

In Reference~\onlinecite{KieneISSCC2021}, In Reference~\onlinecite{KieneISSCC2021}, a successive approximation register (SAR) ADC is developed with the system structure depicted in Fig.~\ref{fig7}(e). The core of this architecture comprises two time-interleaved slices based on an asynchronous loop-unrolled SAR design~\cite{JiangJSSC2012}, which exhibits robustness to temperature variations. The utilization of asynchronous operation and unrolled loops enables power savings and reduction in critical path delay, respectively. Each slice's comparator can be calibrated using two additional input pairs generated by 6-bit resistor-ladder calibration DACs (CDAC). A clock receiver is employed to shape the clock signal for managing the sampling conversion processes. Characterization results at 4K demonstrate that this ADC achieves a 36.2 dB SNDR at 1 GS/s in 6-bit mode, while maintaining an SNDR>44dB below 500 MS/s in 8-bit mode, with a total power consumption of only 10.6mW. Assuming the reflection frequencies of each qubit are uniformly distributed within the Nyquist region with a spacing interval of 10Hz, it would be possible to read out up to 20 qubits using the DAC, resulting in nominal power consumption of only 0.5mW per qubit.

\subsubsection{LO Generattion Module}\label{section_3_3_3}

%\bt{\blue{General statement}}. 
Apart from the control and readout components, another crucial module of Cryo-CMOS ICs is the local oscillator (LO) generation, which provides microwaves for both control and readout modules. Upon delving into the black box of LO generation module, we can identify an oscillator and a PLL~\cite{XinElectronics2023}, with the former typically being a constituent part of the latter (Fig.~\ref{fig2n}). In this section, we present the designs and performance characteristics of both the oscillator and PLL.

\begin{table}[width=0.45\textwidth,pos=!h]
\caption{Performances of cryogenic multiplexers/demultiplexers}
\label{table3}
\begin{tabular*}{\tblwidth}{@{} LLLLLL@{} }
\toprule
~ & \cite{PotoonikQST2021} & \cite{JeongIEDM2022} & \cite{AcharyaVLSI2022} & \cite{HornibrookPRA2015} \\
\midrule
\makecell[l]{Technology} & CMOS & \makecell[l]{InGaAs \\HEMT} & CMOS & \makecell[l]{GaAs \\HEMT} \\ \hline
\makecell[l]{Insertion \\loss (dB)} & 1.6@6GHz & 4.1@4GHz & 2@4GHz & 17@4GHz & ~ \\ \hline
\makecell[l]{Isolation \\(dB)} & 34@6GHz & 45@4GHz & \makecell{40@4GHz \\ 25@8GHz} & 37@4GHz & ~ \\
\bottomrule
\end{tabular*}
\end{table}

%\bt{\blue{CMOS-based works}}. 
In Section~\ref{section_3_3_2}, the VCO design discussed in~\cite{RuffinoISSCC2021} has been previously addressed, thus our focus shifts to Ref.~\onlinecite{GongISSCC2020, PengISSCC2022, ZhangISSCC2023}. The system structure of the VCO presented in~\cite{GongISSCC2020} is illustrated in Fig.~\ref{fig7}(a). In this particular design, a digital calibration loop is incorporated into the oscillator to automatically adjust the differential-mode (DM) and common-mode (CM) capacitor banks. Notably, adjusting the oscillator CM resonance method effectively minimizes both white and 1/f noise sources as demonstrated by previous works such as ~\cite{MurphyISSCC2015, ShahmohammadiISSCC2015}. The objective of the calibration loop is to determine optimal $b_{\rm CD}$ and $b_{\rm CC}$ codes that maximize the 2nd harmonic amplitude $A_{\rm H2}$. A flow chart outlining its procedure can be found in Fig.~\ref{fig7}(b). Initially, set the finite state machine (FSM) to ensure that $V_{\rm DAC} > V_{\rm PD}$ while setting $b_{\rm CD}$ and $b_{\rm CC}$ codes to their maximum and minimum values respectively. Calibration proceeds by incrementing/decrementing these codes by 2/1 each time (to maintain nearly constant total capacitance $C_{\rm C}+C_{\rm D}$ and oscillation frequency ($F_o$)) until reaching a point where $V_{\rm CP}=V_{\rm DD}$. However, if maximum/minimum values for both $b_{\rm CC}$ and $b_{\rm CD}$ are reached before achieving final output conditions, decrement $D_{\rm DAC}$ code by 1 while resetting $b_{\rm CC}$ and $b_{\rm CD}$ back to their initial states. Reference~\onlinecite{ZhangISSCC2023} presents a calibration-free VCO with a tuning range of 12.8-to-16.5 GHz.The main structure adopts class-F architecture with a 1:2 transformer while incorporating an eight-shape head resonator.This design enables dual-peak CM second-harmonic resonance alignment along with DM third-harmonic resonance alignment,resulting in significant reduction of 1/f corner.

%\bt{\blue{SiGe BiCMOS based works}}. 
The VCO, based on SiGe BiCMOS technology as described in~\cite{PengISSCC2022}, exhibits a smaller 1/f corner compared to conventional CMOS technology. The system structure is illustrated in Fig.~\ref{fig7}(d). This mode-switchable structure enables the left and right VCOs to operate at tunable frequencies of 13-to-15.5 GHz and 15-to-17.5 GHz, respectively.

%\bt{\blue{Performance summary}}. 
Finally, we evaluate the performance of the aforementioned VCOs based on three key metrics: phase noise (PN), figure of merit (FoM), and 1/f corner frequency. PN is defined as the ratio between the power attenuation in a unit Hz ($P_0$) and the total power ($P$), typically expressed as $PN=10\log(P_0/P)$. FoM is closely related to PN, while 1/f corner refers to the frequency at which flicker noise and thermal noise amplitudes are equal. In~\cite{GongISSCC2020, PengISSCC2022, ZhangISSCC2023}, comparable PNs and FoMs were achieved. However,~\cite{PengISSCC2022} demonstrated the lowest 165-490 kHz 1/f corner at cryogenic temperature, whereas~\cite{ZhangISSCC2023} exhibited superior performance in terms of 1/f corner compared to~\cite{GongISSCC2020}. Additionally, reference~\onlinecite{GongJSSC2023} presents an integer-N DAPLL that incorporates a dynamic-amplifier-based charge-domain subsampling phase detector (PD) for high phase-detection gain. This PLL achieves -78.4 dBc reference spur level, 75fs RMS jitter, and consumes only 4 mW of power; however, detailed design aspects will not be discussed in this paper.

\subsubsection{Multiplexer and Demultiplexer}\label{section_3_3_4}

%\bt{\blue{General statement}}. 
This section focuses on the multiplexer and demultiplexer, as depicted in the proposed structure (Fig.~\ref{fig4}). In practical implementations, these two blocks are not integrated with the modulator and RF receiver. However, they play a crucial role in significantly mitigating interconnection complexity and may serve as the first CMOS electronics to be deployed at millikelvin temperatures. We provide a comprehensive review of relevant works from Refs.~\onlinecite{PotoonikQST2021, AcharyaVLSI2022}, along with studies utilizing nanowires~\cite{Steins2023, SodergrenNL2022} and HEMT technology~\cite{JeongIEDM2022}.

\begin{figure*}
\centering
\includegraphics[width=1\linewidth]{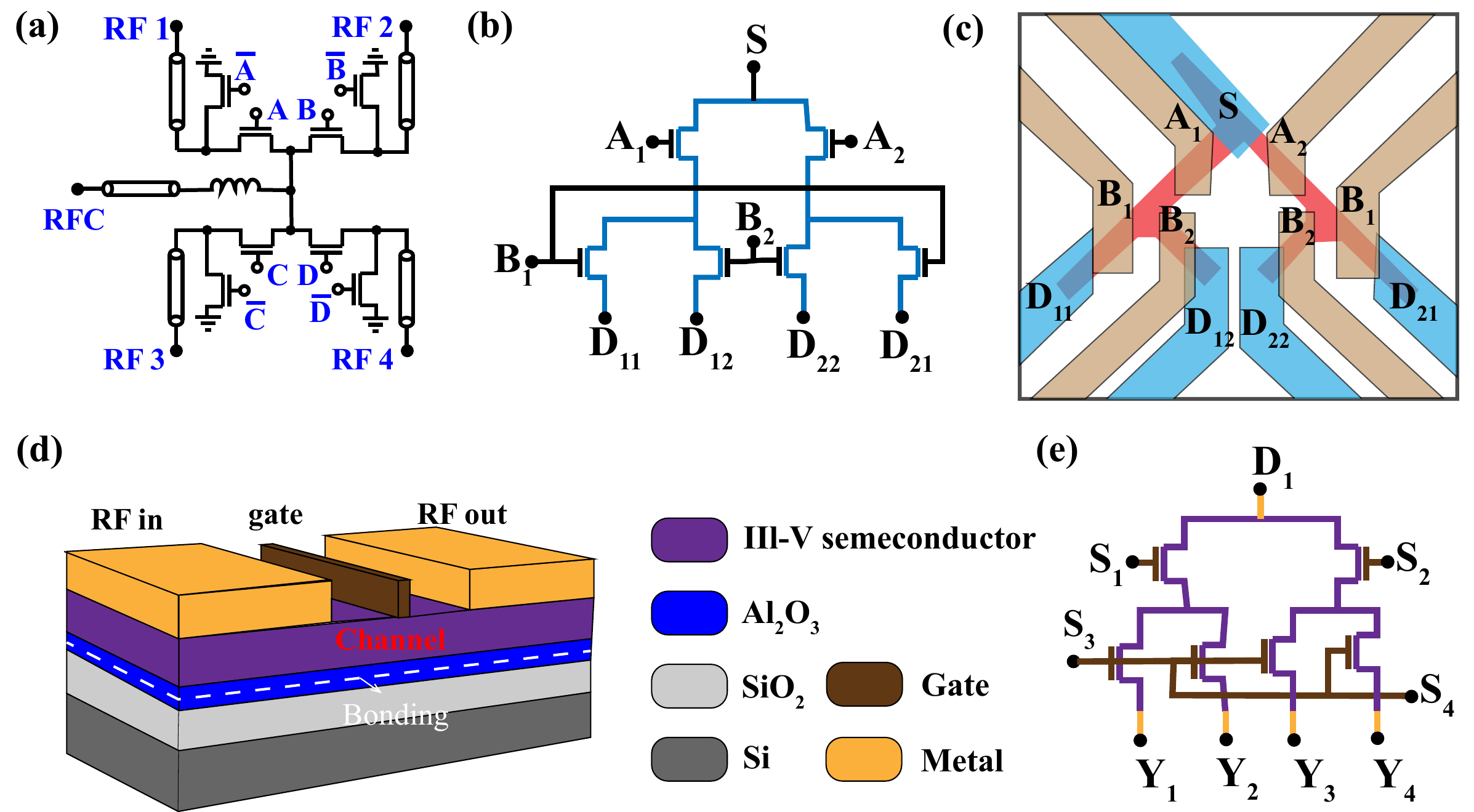}
\caption{\bt{Schematic diagram and layout of multiplexers.} (a) Schematic diagram of CMOS multiplexer in \cite{PotoonikQST2021}. (b) Schematic diagram of CMOS multiplexer in \cite{SodergrenNL2022}. (c) Layout of CMOS multiplexer in \cite{SodergrenNL2022}. (d) Geometric structure of HEMT in \cite{JeongIEDM2022}. (e) Schematic diagram of HEMT-based multiplexer in \cite{JeongIEDM2022}. Panel (a) is from Ref.~\cite{PotoonikQST2021}, (b) and (c) are from Ref.~\cite{SodergrenNL2022}, (d) and (e) are from Ref.~\cite{JeongIEDM2022}. }\label{fig6}
\end{figure*}

%\bt{\blue{CMOS-based works}}. 
A single-pole-4-throw (SP4T) reflective CMOS multiplexer based on TSMC 28-nm CMOS technology is implemented in Ref.~\onlinecite{PotoonikQST2021}, as depicted in Fig.~\ref{fig6}(a). The NMOS transistors' length and width are optimized to minimize insertion loss and enhance isolation performance. Specifically, this design achieves an 'on'-state resistance $R_{\rm on}$ below $10\Omega$ (significantly lower than the transmission line impedance of $50\Omega$), along with a low 'off'-state capacitance $C_{off}<50 fF$. Additionally, a 450pH inductor is incorporated at the common RF port (RFC), while the control logic core and switch transistor are safeguarded by the input-output ring structure. To improve isolation, a series-shunt switch topology~\cite{LiIEEE2010} is adopted. Building upon~\cite{PotoonikQST2021}, ~\cite{AcharyaVLSI2022} further enhances the design by introducing two additional switches for each output RF port, which terminate the transmission lines connected to the quantum chip when these switches are turned off.

%\bt{\blue{mHEMT and nanowire-based works}}. 
The multiplexer, composed of InGaAs quantum well 2DEG-based switches, is developed in Ref.~\onlinecite{JeongIEDM2022}. The device structure of the 2DEG RF switch is illustrated in Fig.~\ref{fig6}(d). When the control voltage $V_{\rm gate}$ is set to zero, the RF signals can pass through the input CPW (coplanar waveguide) line, 2DEG layer, and output CPW line. However, when $V_{\rm gate} = -350 mV$, depletion occurs in the 2DEG layer leading to switch-off state. Combining this basic structure with additional components, a schematic of the multiplexer proposed in~\cite{JeongIEDM2022} is presented in Fig.~\ref{fig6}(e). Furthermore, we introduce designs based on nanowires from references~\cite{Steins2023,SodergrenNL2022}. Both the schematic and layout of a 1$\times$4 multiplexer are depicted in Fig.~\ref{fig6}(b) and Fig.~\ref{fig6}(c), respectively for reference ~\cite{Steins2023}. Similar to the structure described in ~\cite {JeongIEDM2022}, only difference lies in replacing gates, drains or sources along with conducting channels by nanowires. In Ref.~\onlinecite{SodergrenNL2022}, an even larger multiplexer capable of addressing 512 individual quantum devices (qubits or others) using just 37 control lines has been demonstrated; however it employs a crossbar structure for gating which differs slightly from that used by Steins et al.~\cite{Steins2023}.

%\bt{\blue{Performance summary}}. 
After reviewing all the multiplexers, we present the performance summary in Tab.~\ref{table3}, which includes two key metrics: insertion loss and isolation. The meaning of the first metric is self-explanatory, while isolation refers to the signal power attenuation of the enabled channel compared to the disabled channels. Among all works listed in the table, reference~\cite{AcharyaVLSI2022} achieves a remarkable insertion loss of only 2 dB, whereas~\cite{JeongIEDM2022} demonstrates an impressive isolation of 45 dB between channels. It should be noted that both articles based on nanowires do not address RF signals and lack RF characterization results. However, due to their compact size advantage, these schemes may exhibit significantly lower power dissipation compared to other approaches.

\subsubsection{Performances Summary}\label{section_3_3_5}

%\bt{\blue{Prelude}}. 
In Sections~\ref{section_3_3_1} and ~\ref{section_3_3_2}, we have divided the entire integrated circuits (ICs) into modules and blocks for ease of introduction. However, a comprehensive introduction to the performances of both the entire ICs and each individual module has not been provided. In this section, we present an overview of the ICs' performances along with their characterization results using qubits (as shown in Tab.~\ref{table4},~\ref{table5}, and~\ref{table6}). Detailed discussions are presented subsequently.

%\bt{\blue{Basic Infromation of Chips}}. 
The initial four lines present essential information regarding the chips, including the qubit platform and CMOS tech nodes. Most of the chips have adopted bulk CMOS tech nodes due to their enhanced stability at cryogenic temperatures. However, FinFET technology is still accepted in~\cite{PatraISSCC2020,XueNature2021,FrankISSCC2022} for area optimization purposes. Additionally, digital direct synthesis employed in~\cite{FrankISSCC2022} renders the circuit less reliable compared to analog circuits. The functions of ICs are provided below the basic information. Fully integrated chips are developed in~\cite{ParkJSSC2021,KangISSCC2022}, but the power consumption of~\cite{ParkJSSC2021} may exceed what can be accommodated by the cryostat. Relatively lower power consumption is achieved in~\cite{KangISSCC2022}, although it prevents achieving the DRAG scheme due to sharing a pulse shaping code between both I and Q channels. Moreover, there are doubts about whether the pulse shaping method adopted in~\cite{KangISSCC2022} provides sufficient resolution or not. As Google's latest work, channels presented in ~\cite{YooISSCC2023} enable control over two tunable transmons coupled by a coupler. Meanwhile, IBM's work ~\cite{FrankISSCC2022} does not integrate DC bias blocks since they follow a technical roadmap that adopts fixed-frequency transmon and cross-resonance gate techniques ~\cite{RigettiPRB2010}.

\begin{figure}
\centering
\includegraphics[width=1\linewidth]{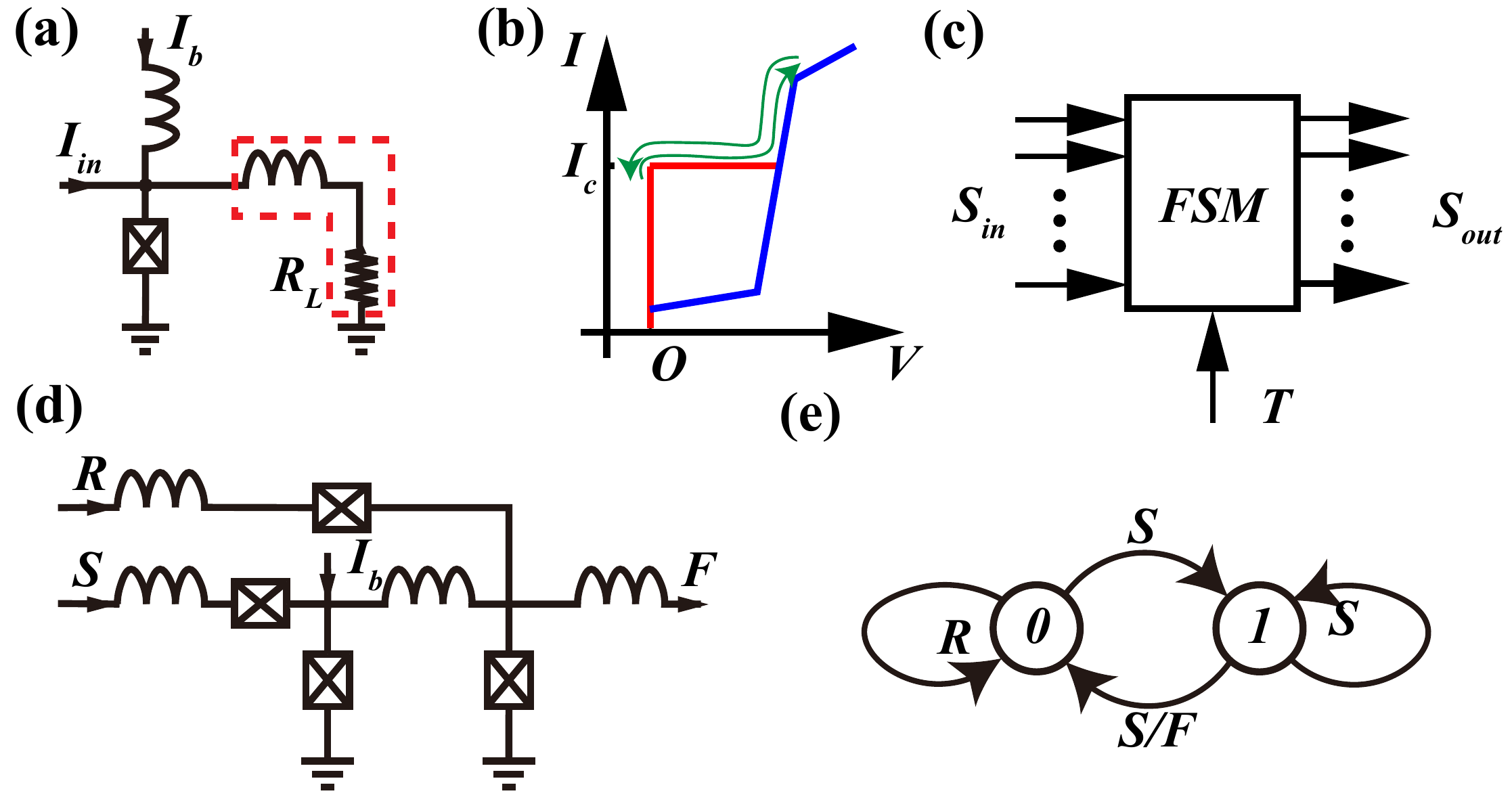}
\caption{\bt{Principles of RSFQ circuits.} (a) Basic circuit unit in RSFQ tech. (b) Current-voltage relation of Josephson junction. (c) Finite state machine. (d) RS flip flop circuit. (e) State transition diagram of RS flip flop. }\label{fig8}
\end{figure}

%\bt{\blue{Function and characterlization results of chips}}. 
The performances of these integrated circuits (ICs) are categorized into three sections in Tab.~\ref{table4},~\ref{table5},~\ref{table6}. Our focus here is on frequency-division multiplexing (FDM) and its characterization results. In terms of XY control, the work by Patra et al.~\cite{PatraISSCC2020}, Xue et al.~\cite{XueNature2021}, and Park et al.~\cite{ParkJSSC2021} achieved a nominal 32 qubits per channel FDM due to the presence of 16 numerically controlled oscillators (NCOs) before the digital-to-analog converter (DAC). However, it should be noted that the coherent time of qubits may deteriorate due to the complex electromagnetic background associated with FDM, thus necessitating further investigation to verify the reliability of FDM control. Unlike FDM control, FDM readout has gained wide acceptance in both superconducting and semiconductor platforms as reported in previous works such as Hornibrook et al.'s study~\cite{HornibrookAPL2014}. ICs equipped with integrated radio-frequency receivers all support FDM, achieving a nominal 16 qubits per channel as demonstrated by Kang et al.'s work~\cite{KangISSCC2022}. The upper limit for FDM depends on factors such as frequency spacing, bandwidth, and analog-to-digital converter (ADC) sampling rate. While most ICs have undergone qubit tests, majority of control tests remain limited to simple Rabbi Oscillation experiments conducted for feasibility verification purposes only; even fewer studies have been dedicated to readout tests compared to control ones. Nevertheless, recent works by IBM's Frank et al.~\cite{FrankISSCC2022} and Google's Yoo et al. ~\cite{YooISSCC2023} have successfully benchmarked quantum logic gates performance where their error rates are comparable or sometimes even better than those achieved using state-of-the-art room temperature electronics.

\section{Rapid Single Flux Quantum Technology}\label{section_4}

%\bt{\blue{General statement}}. 
This section provides a concise overview of the RSFQ technology, which is structured as follows. The first part introduces the historical background of RSFQ, fundamental concepts related to SFQ, RS flip-flop circuitry, and a general FSM model. The second part focuses on qubit control methods utilizing sequences of SFQ pulses instead of microwaves. Additionally, it discusses the potential for developing digital ASICs based on RFSQ technology.

\subsection{ Basic Concepts and Circuits}\label{section_4_1}

%\bt{\blue{History of RSFQ tech}}. 
Similar to the superconducting quantum computing platform, the Josephson junction also serves as the core component of RSFQ circuits. The fundamental unit of an RSFQ circuit is depicted in Fig.~\ref{fig8}(a). Under normal conditions, the current bias $I_{\rm b}$ is maintained below the critical current $I_{\rm c}$ of the Josephson junction. However, when a nonzero input current $I_{\rm in}$ enters the unit, it results in a total current ($I_{\rm b}+I_{\rm in}$) that exceeds $I_{\rm c}$ and induces non-negligible resistance deviating from its superconducting state ($I<I_{\rm c}$)~\cite{GheewalaIEEE1982}. This effect can be exploited to construct a digital circuit resembling conventional semiconductor counterparts. Initially pursued by IBM during the 1970s or early 1980s, their project employed underdamped Josephson junctions exhibiting hysteresis in DC I-V characteristics~\cite{Anacker1980}, as shown in Fig.~\ref{fig8}(b). Notably, this high and low voltage-based bit representation could only achieve clock frequencies up to 1 GHz~\cite{LikharevIEEE1991}, comparable or even slower than its competitors at that time (GaAs logic and CMOS technology). Consequently, this project was discontinued. Nevertheless, pulse-based RSFQ emerged subsequently based on IBM's efforts and modern RSFQ matured through research endeavors during late 1980s~\cite{Likharev1986,MukhanovIEEE1987}. In this advanced technology variant, overdamped Josephson junctions are utilized with "0" and "1" encoded respectively by presence or absence of a single flux quantum.

%\bt{\blue{Concept and Circuit unit of RSFQ}}. 
The single flux quantum (SFQ) is a fundamental concept in physics, representing the minimum magnetic flux $\Phi_0=h/2e$. In RSFQ technology, SFQs propagate along the circuit. To enhance comprehension, we revisit the circuit shown in Fig.~\ref{fig8}(a), assuming $I_b\sim I_c$, which leads to a transient response of the Josephson junction depicted by the trajectory in Fig.~\ref{fig8}(b). This figure illustrates that the Josephson junction transitions from a non-zero resistance state back to its superconducting state rapidly, known as an overdamping process mentioned earlier. From a physical perspective, this transition involves a $2\pi$ phase leap in the superconducting phase of the Josephson junction and generates a single SFQ. However, due to the open topology of the circuit shown in Fig.~\ref{fig8}(a), this SFQ cannot be stored but propagates instead. By considering the effective impedance of the transmission line ($R_{\rm L}$ in Fig.~\ref{fig8}(a)), we can equivalently describe this propagating SFQ as a sharp voltage pulse $V(t)$ (in state-of-the-art Nb-based processing technology, such pulses have an amplitude of 2 mV and width of 1 ps~[X]). According to Faraday law $\dot{\Phi}=V$, $V(t)$ satisfies
\begin{equation}
\Phi_0 = \int_{\rm -\infty}^{\rm +\infty} V(t)dt.
\end{equation}

%\bt{\blue{RSFQ-based RS flip flop}}. 
The RS flip flop circuit is illustrated in Fig.~\ref{fig8}(d). This circuit comprises a DC superconducting quantum interferometer (SQUID) with Josephson junctions of similar size ($I_{\rm c3}\approx I_{\rm c4}\approx I_{\rm c}$), and a current bias $I_{\rm b}=0.8 I_{\rm c}$. Two stable states ("0" and "1") can exist in the loop, where the "0" and "1" states correspond physically to additional persistently counterclockwise and clockwise currents $I_{\rm p}\approx \Phi_0/2L$, respectively. When the loop is in the "0" state, the total current through J3 is approximately $I=I_{\rm b}/2+I_{\rm p}\approx I_{\rm c}$. Subsequently, an SFQ pulse or smaller signal arriving at S causes J3 to undergo a $2\pi$ phase leap, setting the loop state to a clockwise "1". Similarly, if the "1" state is maintained in the loop, an incoming pulse at R resets the state from "1" to "0", generating an SFQ pulse at F. Incorrect operations are rejected by J1 and J2. For instance, if a pulse is received at S while the loop is in the "1" state, it will cause J2 to switch ($2\pi$ phase leap) instead of J3. Overall, this circuit realizes destructive read-out functionality of an RS flip flop with its corresponding state transition diagram shown in Fig.~\ref{fig8}(e). Nondestructive readout and combinational logic can also be achieved using RSFQ technology; for further details on this topic refer to~\cite{LikharevIEEE1991}.

%\bt{\blue{RSFQ-based finite state machine}}. 
An abstract model has been developed in~\cite{LikharevIEEE1991} to describe an FSM from a computer science perspective, where $S_{\rm in}$ and $S_{\rm out}$ represent the input and output ports respectively, and the clock $T$ denotes SFQ pulse sequence evenly spaced in time. For a single-bit input or output port, "0" and "1" are represented by the presence or absence of an SFQ pulse within a clock period, irrespective of arrival time or pulse count. At the end of each period, the FSM state changes and $S_{\rm out}$ is updated (generating an SFQ pulse for a result of "1", while remaining unchanged for a result of "0"). Due to the narrow-width nature of SFQ pulses, RSFQ technology can achieve clock frequencies beyond 300 GHz with ultra-low power dissipation~\cite{LikharevIEEE1991}.

\subsection{Controlling Qubits Through SFQ Pulses}\label{section_4_2}

%\bt{\blue{Related studies of SFQ-pulse-based qubit control}}. 
The coherent manipulation of superconducting qubits using a sequence of SFQ pulses was initially introduced by R. McDermott et al. in~\cite{McDermottPRA2014}. This study examines the theoretical foundations and simulated fidelity of the Y gate. Subsequent research on XY control includes experimental test results~\cite{LeonardPRA2019} and optimization of gate fidelity~\cite{LiebermannPRA2016, LiPRA2019}. Other works, such as 2-qubit control~\cite{JokarQCE2021} and active Z gate implementation using SQF/RSFQ technology~\cite{WangPRA2023}, have also been conducted in recent years. However, due to our limited knowledge, we will only review the simplified theoretical model of this control scheme.

%\bt{\blue{Delta function approximation of SFQ pulse}}. 
The control sequence $s(t)$, initially proposed in~\cite{McDermottPRA2014}, corresponds to the clock signal depicted in Fig.~\ref{fig8}(c). It can be mathematically represented as $s(t)=\sum_{n=0}^{\rm N-1} V(t-nT)$, where N denotes the number of pulses and T represents the time spacing between adjacent pulses. Considering that the width of SFQ is significantly shorter (around 1-2 ps) than the Larmor period (approximately 200 ps), it is reasonable to approximate the SFQ pulse ($V(t)$) as a Dirac delta function ($\Phi_0\delta (t)$)~\cite{McDermottPRA2014}. Consequently, we can rewrite the expression for the pulse sequence as follows
\begin{equation}
s(t) = \Phi_0 \sum_{\rm n=0}^{\rm N-1} \delta(t-nT),
\end{equation}
and its Fourier transportation $F_{\rm s(t)}(\omega)= \int_{-\infty}^{\infty} s(t) e^{-i\omega t}$ could be expressed as
\begin{equation}
F_{\rm s(t)}(\omega) = \Phi_{0} \frac{\sin(n\omega_0 T/2)}{\sin(\omega_0 T/2)}.
\end{equation}
If we model the transmon qubit as an ideal LC resonator and couple it with SFQ pulses via capacitor $C_{\rm c}$ (as shown in Fig.~\ref{fig1}(b)), the energy stored in the resonator ($E$) can be expressed as follows~\cite{McDermottPRA2014}
\begin{equation}
E = \frac{\omega_0^2 C_{\rm c}^2}{2(C_{\rm c} + C)}|F(\omega)|^2,
\end{equation}
However, due to the anharmonicity of transmon and spectral broadening of the control signal, the number typically exhibits a larger value in practical transmon control experiments. Other non-ideal effects, such as the non-zero width of the SFQ pulse and sequence jitter, also contribute to the degradation of coherent control performance. For ease of comprehension, we focus on deriving the evolution operator for this control process while considering only leakage from $|1\rangle$ to $|2\rangle$.

%\bt{\blue{Coupling between the SFQ pulse and transmon}}. 
The total Hamiltonian of this system, within the framework of weak capacitive coupling approximation, can be expressed as~\cite{KrantzAPR2019}
\begin{equation}
\hat{H}_{\rm total}= \hat{Q}^2/2C^{\prime}-E_{\rm J}\cos\hat{\phi} + (C_{\rm c}/C^{\prime})s(t)\hat{Q} = \hat{H}_0 + \hat{H}_{\rm int},
\end{equation}
where $C^{\prime}=C+C_c$, and $\hat{H}_0$ and $\hat{H}_{int}$ are free and interaction Hamiltonian respectively. 
We denote the eigenstates and eigenvalues of $\hat{H}_0$ as $\left\{|n\rangle | n=0,1,2\cdots \right\}$ and $\left\{E_n| n=0,1,2,\cdots \right\}$ respectively. Consequently, we can express the Hamiltonian $\hat{H}_0$ as a summation over all $n$, given by $\hat{H}_0=\sum_{n=0}^{\infty} E_n |n\rangle\langle n|$. By doing so, we are able to derive the free evolution operator
\begin{equation}
\hat{U}_{0}(t)=\exp{\left(-\frac{i}{\hbar} \sum_{n=0}^{\infty}(E_n |n\rangle \langle n|) t\right)}.
\end{equation}
With this equation, further development is straightforward. However, the interaction picture~\cite{JJSakurai2020} is hard to use due to the delta function in SFQ pulse and expansion complexity of $\hat{U}^{\dagger}_0 \hat{H}_{int} \hat{U}_0$. So we introduce the method in~\cite{LeonardPRA2019} which derives the total evolution operator in the Schr\"{o}dinger picture. According to~\cite{LiPRA2019}, evolution operation of a single SFQ pulse $\hat{U}_{SFQ}(t)$ can be expressed as
\begin{equation}
\hat{U}_{\rm SFQ}= \exp{\left(-\frac{i C_{\rm c}}{\hbar C^{\prime}}\int_{0^-}^{t}\Phi_0 \delta(t) dt\hat{Q}\right)} = \exp{\left(-\frac{i C_{\rm c} \Phi_0}{\hbar C^{\prime}} \hat{Q}\right)}.
\end{equation}
The effect of $\hat{H}_0$ is disregarded during the interaction due to the nature of the delta function, and $\hat{U}_{\rm SFQ}$ remains unaffected by the initial and final time of interaction. Utilizing the derived evolution operators, the complete evolution operator can be expressed analytically using the time-order operator $\mathcal{T}$, which reads
\begin{equation}
\hat{U}_{\rm s(t)} = \mathcal{T}(\hat{U}_0(T)\hat{U}_{\rm SFQ}\hat{U}_0(T)\cdots \hat{U}_0(T)\hat{U}_{\rm SFQ}).
\end{equation}
In order to account for the leakage mentioned in the previous paragraph, we consider the subspace spanned by states $\left\{ |0\rangle , |1\rangle , |2\rangle \right\}$. Subsequently, we derive the matrix representations of the two evolution operators (Equation (29) and (30)) as follows
\begin{equation}
\hat{U}_0(t) =\begin{bmatrix}1&0&0\\ 0&e^{-i\omega_{\rm q} t}&0\\ 0&0&e^{-i\omega_{\rm q} (2-\eta) t} \\ \end{bmatrix}
\end{equation}
\begin{equation}
\begin{aligned}
&\hat{U}_{\rm SFQ} =\\ &\frac{1}{\kappa^2}\begin{bmatrix}\lambda^2 + \cos(\kappa\delta\theta/2) & -\kappa\sin(\kappa\delta\theta/2) & 2\lambda\sin^2(\kappa\delta\theta/4)\\ \kappa\sin(\kappa\delta\theta/2) & \kappa^2 \cos(\kappa\delta\theta/2) & -\kappa \lambda \sin(\kappa\delta\theta/2)\\ 2\lambda \sin^2(\kappa\delta\theta/4) & \kappa\lambda \sin(\kappa\delta\theta/2) & 1+\lambda^2 \cos(\kappa\delta\theta/2)\end{bmatrix}
\end{aligned},
\end{equation}
where the $\omega_{\rm q}\equiv \omega_{10}$ means the qubit frequency, $\eta=1-\omega_{21}/\omega_{\rm q}$ is the anharmonicity of transmon, $\delta \theta=(2\Phi_0/\hbar)(C_c/C^{\prime})\langle 1|\hat{Q}|0\rangle$, $\lambda =\langle 2| \hat{Q} |1\rangle / \langle 1|\hat{Q}|0\rangle$ and $\kappa = \sqrt{\lambda^2 + 1}$. 
Assuming the transmon behaves as an ideal two-level system ($\lambda = 0$), the impact of $U_{SFQ}$ is a simple rotation of the qubit state vector $\delta\theta$ along the y-axis on the Bloch sphere, which is equivalent to applying gate $Y_{\delta\theta}$. Considering the overall evolution driven by a sequence of SQF pulses (s(t)), under the condition $2\pi/\omega_{\rm q}=T$, non-ideal leakage can be suppressed. However, it has been shown in~\cite{McDermottPRA2014} that using a symmetrical sequence may not yield optimal results; instead, adjusting the intervals between SFQ pulses allows for achieving fidelity exceeding 99.9\%~\cite{LiebermannPRA2016}. Moreover, in~\cite{LiPRA2019}, utilizing these equations and numerical simulations enabled high-fidelity control (over 99.99\%) of qubits resonating at 20 distinct frequencies within a single global RSFQ clock~\cite{LiPRA2019}.

%\bt{\blue{Summary of the RSFQ-based control and readout}}. 
In this section, we focus on the potential of developing Cryo ICs based on RSFQ technology. Firstly, a notable advantage of the RSFQ control scheme is its compatibility with the superconducting platform, as both the superconducting quantum processor and RSFQ technology rely on Josephson junctions. Secondly, the lower power consumption of RSFQ logic enables deployment of Cryo ICs even at millikelvin stages. Thirdly, the exceptional clock frequency offers ultra-fast feedback for QEC applications. By integrating all three advantages, it becomes possible to integrate C/R circuits on a single chip for future fault-tolerant quantum computing where the readout circuit can also be controlled using RSFQ technology~\cite{McDermottPRA2014} in conjunction with the read-out circuit proposed by Opremcak et al.~\cite{OpremcakPRX2021}. The corresponding system architecture has been proposed by Mukhanov et al.~\cite{MukhanovIEDM2019}. However, there are still two deficiencies: 1. The control fidelity is lower compared to state-of-the-art RF control systems; for instance, an experiment conducted in 2019 achieved only 95\% fidelity~\cite{LeonardPRA2019}, which falls significantly short of state-of-the-art 99.9\% fidelity levels; and 2. Microprocessors based on RSFQ technology have larger sizes compared to those based on mature CMOS technology.

\section{Summary and Outlook}\label{section_5}

%\bt{\blue{Summary of challenges of Cryogenic ICs}}. 
In this review, we have comprehensively discussed Cryo-CMOS ICs and RSFQ technology, elucidating their underlying principles, architectural features, and circuit design considerations. However, it is important to acknowledge that the realization of quantum error correction (QEC) remains a significant challenge for most prototype ICs, including those based on RSFQ technology. We identify three key challenges that necessitate attention: power consumption optimization, enhancement of reliability aspects, and resolution of multiplexing issues.
\begin{enumerate}[1)]
\item The power consumption poses the greatest challenge due to the limited cooling power of current refrigerators. At the 4 K stage, the cooling power is approximately 1 W~\cite{Batey2014}, and even lower at the millikelvin stage. Despite advancements in power-saving Cryo-CMOS IC prototypes, each XY control channel still consumes around 10 mW, while Read-out channels consume a few hundred mW (Section~\ref{section_3_3_5}). These levels remain insufficient for fault-tolerant systems, particularly considering the escalating number of LO signals that contribute to overall power consumption.
\item The reliability of Cryo-CMOS FETs is primarily affected by the modeling problem associated with FETs. Despite extensive research efforts, there remains a dearth of mature commercial foundry-based models for Cryo-CMOS FETs, and the presence of unpredictable noise can exacerbate these issues. While Cryo-ICs can achieve fidelity comparable to room temperature electronics~\cite{FrankISSCC2022, YooISSCC2023}, it becomes imperative to consider additional challenges such as crosstalk between channels as the number of channels increases.
\item The challenge of multiplexing arises from the tradeoff between interconnection complexity and qubit decoherence. On one hand, the utilization of multiplexers or frequency-division multiplexing schemes is imperative to minimize cable count. Conversely, these devices and schemes can induce qubit decoherence. Furthermore, it is essential to investigate the feasibility of FDM schemes for control signals, particularly XY control, as well as the impact of non-zero on-off ratios caused by RF front-end (pulse modulator) and multiplexer.
\end{enumerate}

%\bt{\blue{Outlook}}. 
Finally, two individual perspectives on the future advancement of Cryogenic Integrated Circuits (Cryo ICs) are presented.

\begin{enumerate}[1)]
\item Co-integrated digital circuits should be considered as the future of Cryogenic C/R ICs. In~\cite{ZettlesISSCC2022}, a distributed control system for quantum computing is proposed, which offers enhanced flexibility in executing the QEC algorithm due to its low-delay feedback control. If feasible, we propose solidifying the QEC algorithms within the digital component of Cryo ICs to minimize power consumption and latency (similar to FPGA in electronic computers). To some extent, RSFQ-based digital circuits exhibit greater potential than CMOS ones but are constrained by their integration level (Section~\ref{section_4_2}). As an alternative approach, we suggest employing RSFQ technology for constructing the active control segment of QEC during initial stages while retaining CMOS technology for other components.
\item The integration of quantum processor design and Cryo ICs simulation should be pursued, if feasible. This perspective is rooted in coherence considerations, as the intricate electromagnetic characteristics of control signals, cables, and chips themselves can all undermine qubit coherence. Currently, there are already relevant proposals (~\cite{RaduVLSI2021, vanDiRPA2019}), and future research should concentrate on establishing an integrated simulation and design platform akin to commercial EDA tools used in classical IC design. Once the technology reaches maturity, co-integrated chips comprising qubits alongside control and readout circuits may become viable.
\end{enumerate}

\section*{Appendix}

This appendix presents three tables compiling the performance data of Cryo-CMOS prototype chips.

\begin{table*}[width=1\textwidth,pos=!h]
\caption{Performances of cryogenic multiplexers/demultiplexers}
\label{table4}
\begin{tabular*}{\tblwidth}{@{} |L|L|L|L|L|@{} }
\hline
~ & ISSCC 2021 \cite{RuffinoISSCC2021} & ISSCC 2021 \cite{PrabowoISSCC2021} & ISSCC 2020 \cite{PatraISSCC2020} & Nature 2021 \cite{XueNature2021}\\
\hline
\makecell[l]{Qubit Platform} & Quantum dots & Spin qubits & Spin qubits/Transmons & Spin qubits\\ \hline
\makecell[l]{CMOS Tech Node} & \makecell[l]{40-nm \\bulk CMOS} & \makecell[l]{40-nm \\bulk CMOS} & \makecell[l]{22-nm \\FinFET} & \makecell[l]{22-nm \\ FinFET}\\ \hline
\makecell[l]{Chip Area} & 2.8 $mm^2$ & 0.68 $mm^2$ & 16 $mm^2$ & 16 $mm^2$\\ \hline
\makecell[l]{Temperature} & 3.5 K & 4.2 K & 3 K & 3 K\\ \hline
\makecell[l]{\textbf{Function}} & ~ & ~ & ~ & ~\\ \hline
\makecell[l]{XY Control \\ (Pulse Modulator)} & ~ & ~ & 1 Channel I/Q SSD & 1 Channel I/Q SSD\\ \hline
\makecell[l]{Z/g Control \\ (DC Bias)} & ~ & ~ & ~ & ~\\ \hline
\makecell[l]{Readout \\ (RF Reciever)} & 1 Channel & 1 Channel & ~ & ~ \\ \hline
\makecell[l]{LO \\Generation} & PLL & External & External & External \\ \hline
\makecell[l]{Digital \\control} & ~ & ~ & Instruction set & \makecell[l]{digitally intensive\\ architecture} \\ \hline
\makecell[l]{\textbf{Performances} \\\textbf{of Control}} & ~ & ~ & ~ & ~ \\ \hline
\makecell[l]{Frequency \\ Range} & ~ & ~ & 2-20 GHz & 2-20 GHz\\ \hline
\makecell[l]{XY Pulse \\ Width} & ~ & ~ & Variable & Variable\\ \hline
\makecell[l]{Envelope\\ Waveforms} & ~ & ~ & Arbitrary & Arbitary\\ \hline
\makecell[l]{SFDR} & ~ & ~ & > 45 dB & 46 dB\\ \hline
\makecell[l]{LORR} & ~ & ~ & > 36 dB & 38 dB\\ \hline
\makecell[l]{Sampling \\ frequency} & ~ & ~ & 1 GS/s & ~\\ \hline
\makecell[l]{FDM} & ~ & ~ & \makecell[l]{32 bit/Ch. \\(nominally)} & \makecell[l]{32 (TDM) \\ 2 (FDM)}\\ \hline
\makecell[l]{\textbf{Performances} \\\textbf{of Readout}} & ~ & ~ & ~ & ~\\ \hline
\makecell[l]{Bandwidth} & 5-6.5 GHz & 6-8 GHz & ~ & ~\\ \hline
\makecell[l]{Gain} & 70 dB & 58 dB & ~ & ~\\ \hline
\makecell[l]{IIP3} & >-72 dBm & -50.8 dBm ($\Delta$=50 MHz) & ~ & ~\\ \hline
\makecell[l]{P1dBin} & >-85 dBm & -58.4 dBm & ~ & ~\\ \hline
\makecell[l]{FDM} & ~ & ~ & ~ & ~\\ \hline
\makecell[l]{\textbf{Power} \\ \textbf{consumption}} & ~ & ~ & ~ & ~\\ \hline
\makecell[l]{Total Power \\Consumption} & 108 mW & 70 mW & ~ & 192 mW/qubit\\ \hline
\makecell[l]{XY Control} & ~ & ~ & 1.7 mW/qubit & ~\\ \hline
\makecell[l]{Z/g Control \\(DC Bias)} & ~ & ~ & ~ & ~\\ \hline
\makecell[l]{Readout} & ~ & 70 mW & ~ & ~\\ \hline
\makecell[l]{LO \\Generation} & ~ & ~ & ~ & ~\\ \hline
\makecell[l]{Digital \\Block} & ~ & ~ & 330mW & ~\\ \hline
\makecell[l]{\textbf{Characterization} \\ \textbf{with Qubits}} & ~ & ~ & Rabbi Osc. & \makecell[l]{Rabbi Osc. \\2Q Gate}\\ \hline
\makecell[l]{XY Average \\Gate Error} & ~ & ~ & ~ & ~\\ \hline
\makecell[l]{Z Average \\Gate Error} & ~ & ~ & ~ & ~\\ \hline
\makecell[l]{2 Qubit \\Average Error} & ~ & ~ & ~ & ~\\
\hline
\end{tabular*}
\end{table*}

\begin{table*}[width=1\textwidth,pos=!h]
\caption{Performances of cryogenic multiplexers/demultiplexers}
\label{table5}
\begin{tabular*}{\tblwidth}{@{} |L|L|L|L|L|@{} }
\hline
~ & ISSCC 2021 \cite{ParkISSCC2021} & ISSCC 2022 \cite{FrankISSCC2022} & ISSCC 2023 \cite{YooISSCC2023} & ISSCC 2023 \cite{BardinISSCC2019}\\
\hline
\makecell[l]{Qubit Platform} & Spin qubits & Transmons & Transmons & Transmons\\ \hline
\makecell[l]{CMOS Tech Node} & \makecell[l]{22-nm \\FinFET} & \makecell[l]{14-nm \\FinFET} & \makecell[l]{28-nm \\bulk CMOS} & \makecell[l]{28-nm \\bulk CMOS}\\ \hline
\makecell[l]{Chip Area} & 16 $mm^2$ & 3.2175 $mm^2$ & 1.6 $mm^2$ & 1.6 $mm^2$\\ \hline
\makecell[l]{Temperature} & 3 K & 3 K & 3 K & 3 K \\ \hline
\makecell[l]{\textbf{Function}} & ~ & ~ & ~ & ~\\ \hline
\makecell[l]{XY Control \\ (Pulse Modulator)} & 1 Channel I/Q SSD & 2 Channel I/Q SSD & 2 Channel I/Q Direc. Conv. & 1 Channel I/Q Direc. Conv.\\ \hline
\makecell[l]{Z/g Control \\ (DC Bias)} & ~ & 4 Channels & 3 Channels & ~\\ \hline
\makecell[l]{Readout \\ (RF Reciever)} & ~ & 1 Channel & ~ & ~\\ \hline
\makecell[l]{LO \\Generation} & External & External & External & External\\ \hline
\makecell[l]{Digital \\control} & \makecell[l]{digitally intensive\\ architecture} & Microprocessor & Instruction Sqeuncer & Shift Registers\\ \hline
\makecell[l]{\textbf{Performances} \\\textbf{of Control}} & ~ & ~ & ~ & ~ \\ \hline
\makecell[l]{Frequency \\ Range} & 2-20 GHz & 4.5-5.5 GHz & 4-8 GHz & 4-8 GHz \\ \hline
\makecell[l]{XY Pulse \\ Width} & Variable & Variable & Fixed (22 $T_{clock}$) & Fixed (22 $T_{clock}$) \\ \hline
\makecell[l]{Envelope\\ Waveforms} & Arbitrary & 16 Envelopes & 16 Envelopes & Arbitary\\ \hline
\makecell[l]{SFDR} & 45 dB & ~ & ~ & ~\\ \hline
\makecell[l]{LORR} & 36 dB & ~ & ~ & ~\\ \hline
\makecell[l]{Sampling \\ frequency} & up to 2.5 GHz & 1 GHz & ~ & 1 GHz\\ \hline
\makecell[l]{FDM} & \makecell[l]{32 (TDM) \\ 2 (FDM)} & ~ & ~ & ~\\ \hline
\makecell[l]{\textbf{Performances} \\\textbf{of Readout}} & ~ & ~ & ~ & ~\\ \hline
\makecell[l]{Bandwidth} & \makecell[l]{60-200 MHz \\ (baseband)} & ~ & ~ & ~\\ \hline
\makecell[l]{Gain} & 40-90 dB & ~ & ~ & ~\\ \hline
\makecell[l]{IIP3} & ~ & ~ & ~ & ~\\ \hline
\makecell[l]{P1dBin} & ~ & ~ & ~ & ~\\ \hline
\makecell[l]{FDM} & 6 bit/Ch.& ~ & ~ & ~\\ \hline
\makecell[l]{\textbf{Power} \\ \textbf{consumption}} & ~ & ~ & ~ & ~\\ \hline
\makecell[l]{Total Power \\Consumption} & 384 mW & 23.1 mW/qubit & <4 mW/qubit & < 2 mW/qubit\\ \hline
\makecell[l]{XY Control} & around 30 mW & 54.9\% & ~ & ~ \\ \hline
\makecell[l]{Z/g Control \\(DC Bias)} & ~ & ~ & ~ & ~\\ \hline
\makecell[l]{Readout} & around 20 mW & ~ & ~ & ~\\ \hline
\makecell[l]{LO \\Generation} & ~ & ~ & ~ & ~\\ \hline
\makecell[l]{Digital \\Block} & 330 mW & 45.1\% & ~ & ~\\ \hline
\makecell[l]{\textbf{Characterization} \\ \textbf{with Qubits}} & Rabbi Osi. & Rabbi Osi.& \makecell[l]{Rabbi Osc. \\ 2Q Gate} & Rabbi Osc.\\ \hline
\makecell[l]{XY Average \\Gate Error} & ~ & 0.078\% & 0.17-0.36\% & ~\\ \hline
\makecell[l]{Z Average \\Gate Error} & ~ & ~ & 0.14-0.19\% & ~\\ \hline
\makecell[l]{2 Qubit \\Average Error} & ~ & ~ & 1.2\%$\pm$0.3\% & ~\\
\hline
\end{tabular*}
\end{table*}

\begin{table*}[width=1.05\textwidth,pos=!h]
\caption{Performances of cryogenic multiplexers/demultiplexers}
\label{table6}
\begin{tabular*}{\tblwidth}{@{} |L|L|L|L|L|@{} }
\hline
~ & ISSCC 2023 \cite{KangISSCC2023} & ISSCC 2022 \cite{KangISSCC2022} & VLSI 2021 \cite{KangVLSI2021} & ISSCC 2023 \cite{GuoISSCC2023} \\
\hline
\makecell[l]{Qubit Platform} & Transmons & Transmons & Transmons & Transmons \\ \hline
\makecell[l]{CMOS Tech Node} & \makecell[l]{40-nm \\ bulk CMOS} & \makecell[l]{40-nm \\bulk CMOS} & \makecell[l]{40-nm\\ bulk CMOS} & \makecell[l]{28-nm \\bulk CMOS}\\ \hline
\makecell[l]{Chip Area} & 4.16 $mm^2$ & 5.2 $mm^2$ & 2.34 $mm^2$ & 1.8 $mm^2$\\ \hline
\makecell[l]{Temperature} & 3.5 K & 3.5 K & 77 K & 3.5 K\\ \hline
\makecell[l]{\textbf{Function}} & ~ & ~ & ~ & ~\\ \hline
\makecell[l]{XY Control \\ (Pulse Modulator)} & 4 Channel I/Q Direc. Conv. & 4 Channel I/Q Mod. & 6 Channel I/Q Mod. & 2 Channel Palor Mod.\\ \hline
\makecell[l]{Z/g Control \\ (DC Bias)} & ~ & 4 Channels & ~ & ~\\ \hline
\makecell[l]{Readout \\ (RF Reciever)} & ~ & 2 Channels & ~ & ~\\ \hline
\makecell[l]{LO \\Generation} & LB PLL$\times$2 & LB PLL$\times$2 & ~ & IL-LO\\ \hline
\makecell[l]{Digital \\control} & $\varphi$ and $f$ controller & $\varphi$ and $f$ controller & $\varphi$ and $f$ controller & FIFO\\ \hline
\makecell[l]{\textbf{Performances} \\ \textbf{of Control}} & ~ & ~ & ~ & ~ \\ \hline
\makecell[l]{Frequency \\ Range} & 5.0-6.5 GHz & 4.5-8 GHz & 2-7 GHz & 4-6 GHz \\ \hline
\makecell[l]{XY Pulse \\ Width} & Variable & Variable & Variable & Variable \\ \hline
\makecell[l]{Envelope\\ Waveforms} & \makecell[l]{Cosine Envelopes \\ (DRAG)} & Arbitary (besides DRAG) & Arbitary (besides DRAG) & Arbitary (besides DRAG) \\ \hline
\makecell[l]{SFDR} & ~ & ~ & ~ & ~\\ \hline
\makecell[l]{LORR} & > 41 dB & > 41 dB & > 41 dB & ~\\ \hline
\makecell[l]{Sampling \\ frequency} & 0.7-1.6 GHz & up to 2 GS/s & 2 GS/s & ~ \\ \hline
\makecell[l]{FDM} & ~ & ~ & ~ & ~ \\ \hline
\makecell[l]{\textbf{Performances} \\\textbf{of Readout}} & ~ & ~ & ~ & ~\\ \hline
\makecell[l]{Bandwidth} & ~ & 5-7.25 GHz & ~ & ~ \\ \hline
\makecell[l]{Gain} & ~ & 47 dB & ~ & ~ \\ \hline
\makecell[l]{IIP3} & ~ & > -35 dBm & ~ & ~ \\ \hline
\makecell[l]{P1dBin} & ~ & > -49 dBm & ~ & ~ \\ \hline
\makecell[l]{FDM} & ~ & 16 qubits & ~ & ~ \\ \hline
\makecell[l]{\textbf{Power} \\ \textbf{consumption}} & ~ & ~ & ~ & ~\\ \hline
\makecell[l]{Total Power \\Consumption} & 24.1 mW/qubit & ~ & 5.5 mW/channel & 13.7mW/qubit \\ \hline
\makecell[l]{XY Control} & ~ & 0.49 mW/qubit & 2.9 mW/channel & ~\\ \hline
\makecell[l]{Z/g Control \\(DC Bias)} & ~ & 0-10 mW/channel & ~ & ~ \\ \hline
\makecell[l]{Readout} & ~ & 20mW/channel & ~ & ~ \\ \hline
\makecell[l]{LO \\Generation} & ~ & ~ & ~ & ~ \\ \hline
\makecell[l]{Digital \\Block} & ~ & 3.8 mW/channel & 2.6 mW/channel & ~ \\ \hline
\makecell[l]{\textbf{Characterization} \\ \textbf{with Qubits}} & ~ & Readout & Rabbi Osc. & Rabbi Osc. \\ \hline
\makecell[l]{XY Average \\Gate Error} & ~ & ~ & ~ & ~\\ \hline
\makecell[l]{Z Average \\Gate Error} & ~ & ~ & ~ & ~\\ \hline
\makecell[l]{2 Qubit \\Average Error} & ~ & ~ & ~ & ~\\
\hline
\end{tabular*}
\end{table*}

\bibliographystyle{elsarticle-num.bst}
\bibliography{main.bib}

\end{document}